\documentclass[aps,prx,reprint,superscriptaddress, longbibliography]{revtex4-1}
\usepackage{H1}
\usepackage{amsmath}
\usepackage[pdftex,colorlinks=true]{hyperref}
\hypersetup{
citecolor = blue
}
\usepackage[normalem]{ulem}
\usepackage{amssymb}
\usepackage{amsthm}
\usepackage{amsfonts}
\usepackage{bbm}
\usepackage{array}

\usepackage{scalerel,amssymb}
\def\msquare{\mathord{\scalerel*{\Box}{gX}}}

\begin{document}

\title{Spectroscopic fingerprints of gapped quantum spin liquids, both conventional and fractonic}
\author{Rahul M. Nandkishore}
\affiliation{Department of Physics and Center for Theory of Quantum Matter, University of Colorado, Boulder, CO 80309}
\author{Wonjune Choi}
\author{Yong Baek Kim}
\affiliation{Department of Physics, University of Toronto, Toronto, Ontario M5S 1A7, Canada}

\begin{abstract} 
We explain how gapped quantum spin liquids, both conventional and `fractonic,' may be unambiguously diagnosed experimentally using the technique of multidimensional coherent spectroscopy.  `Conventional' gapped quantum spin liquids (e.g. $Z_2$ spin liquid) do not have clear signatures in linear response, but \textit{do} have clear fingerprints in \textit{non-linear} response, accessible through the already existing experimental technique of two dimensional coherent spectroscopy. Type I fracton phases (e.g. X-cube) are (surprisingly) even easier to distinguish, with strongly suggestive features even in linear response, and  unambiguous signatures in non-linear response. Type II fracton systems, like Haah's code, are most subtle, and may require consideration of high order non-linear response for unambiguous diagnosis. 
\end{abstract}
\maketitle

Quantum spin liquids are phases of quantum matter which feature fractionalized excitations and emergent deconfined gauge fields (for reviews, see \cite{SavaryBalents, KanodaReview}). Recently  an exotic `fractonic' variant of quantum spin liquids has been proposed, which additionally host emergent excitations with restricted mobility (for reviews, see \cite{fractonarcmp, PretkoChenYou}). The most stable and theoretically best understood examples of spin liquids, both conventional and fractonic, have the ground state manifold separated from the rest of the spectrum by an energy gap. The $Z_2$ spin liquid provides a paradigmatic example of a conventional gapped quantum spin liquid, while the X-cube model and Haah code provide paradigmatic examples of gapped fractonic spin liquids, of type I and type II respectively. 

While great strides have been made in the theoretical understanding of gapped quantum spin liquids, these phases have not yet been unambiguously observed in any experimental system. Part of the challenge here is that it is hard to find unambiguous experimental diagnostics for these exotic phases, which are accessible using currently available experimental techniques. This is in contrast to {\it gapless} spin liquids (both conventional and fractonic), which do have clean signatures e.g. in the form of `pinch points' in the dynamical structure factor, which may be probed via neutron scattering \cite{CastelnovoReview, fractonpinch}. 

In this paper we explain how gapped quantum spin liquids, both conventional and fractonic, may be unambiguously diagnosed using already existing spectroscopic tools. Gapped conventional spin liquids do not have clear diagnostics in linear response, but do have unambiguous fingerprints in non-linear response. These may be identified through the technique of two dimensional coherent spectrsocopy (2DCS), originally pioneered in the context of nuclear magnetic resonance and physical chemistry \cite{mukamel, Hamm, Bartholdi, Cundiff}, and recently applied also to solid state systems \cite{mfg, ArmitageWan, choileekim, parameswarangopalakrishnan}. Gapped fracton phases of type-I are actually easier to identify, with strongly suggestive features even in linear response, and unambiguous fingerprints in non-linear response. Fracton phases of type II are most subtle, and may require consideration of high order non-linear response for unambiguous diagnosis.  

The rest of this paper is structured as follows. In Section \ref{sec: models} we introduce the paradigmatic models that we use as examples of gapped spin liquid, type I fracton and type II fracton phases respectively. In Section \ref{sec: lresponse} we explain the properties of the respective phases in linear response, and point out how type I fracton phases (and only type I fracton phases) have clear signatures therein. In Sec.\ref{sec: nlresponse} we explain how to calculate the non-linear susceptibility measured in a 2DCS experiment. In Sec \ref{sec: results} we explain the key signatures of each of the phases considered within 2DCS. We conclude in section \ref{sec: conclusions} with a discussion of outlook and implications. Technical details are relegated to the appendices. 


\section{Phases of interest}
\label{sec: models}
We are interested in three types of phases: conventional gapped spin liquids, type I fractons, and type II fractons. We consider a paradigmatic example of each, as well as a `control' example  which is not a spin liquid. While the model Hamiltonians we write down are for the most part fine tuned to an exactly solvable point, our interest will be specifically in those features of the phase that are robust to perturbations, and which thus do not rely on being at the fine tuned point. The solvable Hamiltonians that we present therefore simply represent `fixed point Hamiltonians' that characterize the respective phases. 

As a non-spin liquid control example, we consider the quantum Ising model in two or three spatial dimensions. This is defined on a two (three) dimensional square (cubic) lattice with spin-$1/2$ variables living on the vertices.
The Hamiltonian is 
\begin{equation}
H = -\frac{1}{2}\sum_{\langle i,j \rangle} Z_i Z_j + \lambda (...)
\end{equation}
where $\langle \cdot, \cdot \rangle$ denotes nearest neighbors, $Z$ and $X$ are Pauli opreators, and $\lambda$ is a parameter controlling strength of perturbations. Here $(...)$ denotes arbitrary local perturbations that fail to commute with the $ZZ$ term but which respect Ising symmetry (the simplest such perturbation being a transverse field, $\sum_k X_k$). Note that in two (three) dimensions, this {\it is} an interacting model in its own right.  We work with periodic boundary conditions. We work mostly in the ferromagnetic phase, $\lambda \ll 1$, when the ground state spontaneously breaks $Z \rightarrow -Z$ symmetry, and the elementary excitations are spin flips. This is not a spin liquid phase and serves as our control example. 
 
As a paradigmatic example of a gapped conventional spin liquid, we consider the perturbed two-dimensional toric code \cite{Kitaev}. This is defined on the two-dimensional square lattice with spin-$1/2$ variables living on the links. The Hamiltonian is 
\begin{equation}
H = - \frac{1}{2}\sum_{v} A_v - \frac{J}{2}\sum_{p} B_p -  \lambda (...)
\end{equation}
where $\sum_v$ indicates a sum over vertices, $A_v$ is a product of four $X$ operators on the four links connected to a vertex $v$, $\sum_p$ indicates a sum over square plaquettes, $B_p$ is a product of four $Z$ operators around a plaquette, $(...)$ denotes arbitrary local perturbations, and  $\lambda \ll1$. The exactly solvable point is $\lambda = 0$. We work with periodic boundary conditions and the ground state which have zero flux penetrating the great circles of the torus. At the exactly solvable point, it has eigenvalue $+1$ under every $A$ and $B$ operators. 

As a paradigmatic example of a type I fracton phase, we consider the (perturbed) X-cube model \cite{Sherrington, fracton2}. This is defined on a three-dimensional cubic lattice with spin-$1/2$ variables living on the links. The Hamiltonian is 
\begin{equation}
H = - \frac12\sum_{v} (A^{xy}_v+A^{yz}_v+A^{zx}_v) - \frac{K}{2} \sum_{c} B_c -  \lambda (...)
\end{equation}
where $\sum_v$ denotes sum over vertices, $A^{ab}_v$ is a product over four $Z$ operators on links connected to vertex $v$ lying in the $ab$ plane, $\sum_c$ denotes sum over cubes, $B_c$ is a product over twelve $X$ operators framing an elementary cube, $\lambda \ll 1$, and $(...)$ denotes arbitrary local perturbations. The exactly solvable point is $\lambda = 0$. We work with periodic boundary conditions and the ground state which have zero flux penetrating the great circles of the torus. At the exactly solvable point it has eigenvalue $+1$ under every $A, B$. The elementary excitations consist of lineons, planeons, and fractons. Lineons are created at the end of a string of X operators and are one dimensional particles (i.e. able to move in only one direction, since the string can change length but cannot change direction without creating additional lineons); planeons are created at the end of a string of Z operators and are two dimensional particles (i.e. able to move in a two dimensional plane). Fractons are created at the corners of a rectangular membrane of Z operators and are totally immobile under local perturbations. All these properties are explained at length in \cite{fracton2, fractonarcmp}. 

As a paradigmatic example of a type II fracton phase, we consider the (perturbed) Haah code \cite{haah}. This is defined on a three dimensional cubic lattice with two spin-$1/2$ variables on every vertex. The Hamiltonian is 
\begin{equation}
H = - \frac12 \sum_{c} A - \frac{\Lambda}{2} \sum_{c} B - \lambda(...)
\end{equation}
where $A$ is a particular product of $X$ type Pauli's and identities around a cube, and $B$ is likewise with $X \rightarrow Z$, $\Lambda \approx 1$, $\lambda \ll 1$, and $(...)$ denotes arbitrary local perturbtions. The exactly solvable point is $\lambda = 0$. The ground state is unique with open boundary conditions (eigenvalue +1 under every A and B at the solvable point). The elementary excitations are \textit{fractons} (totally immobile excitations). There are no subdimensional particles, but a composite of four fractons in a tetrahedral arrangement is locally creatable (and therefore mobile). 

It should be emphasized that while these model Hamiltonians appear exceedingly baroque, nonetheless the {\it phases} are robust. The value of these model Hamiltonians is that they allow for exact calculations. While we will present some exact results obtained at the solvable points ($\lambda = 0)$, our emphasis in what follows will be on {\it universal} features that should be present throughout the phase, and not just at the solvable points. Also, while we have chosen periodic boundary conditions for analytic convenience, experiments of course will have open boundary conditions. However, since the non-linear response we are discussing does not connect distinct topological sectors, the choice of boundary conditions should not make a difference. 

\section{Linear response}
\label{sec: lresponse}
The quantities of interest in linear response spectroscopy are correlators $\langle M^z(t) M^z(0)\rangle$ and $\langle M^x(t) M^x(0)\rangle $ and $\langle M^y(t) M^y(0)\rangle $, where $M^z = \sum Z$ measures the magnetization in the $z$ direction and $M^x$ and $M^y$ are defined analogously. All of these are measurable with the suitable polarization of the incident light. The expectation values are taken on the ground state. Inserting a resolution of the identity in terms of eigenstates of the Hamiltonian, these correlators take the form 
\begin{eqnarray}
\langle M^z(t) M^z(0) \rangle &=& \sum_{\alpha} \langle 0 | \sum Z | \alpha \rangle \langle \alpha | \sum Z | 0 \rangle \exp(i E_{\alpha} t)\nonumber \\
\langle M^x(t) M^x(0) \rangle &=& \sum_{\alpha} \langle 0 | \sum X | \alpha \rangle \langle \alpha | \sum X | 0 \rangle \exp(i E_{\alpha} t) \nonumber\\
\langle M^y(t) M^y(0) \rangle &=& \sum_{\alpha} \langle 0 | \sum Y | \alpha \rangle \langle \alpha | \sum Y | 0 \rangle \exp(i E_{\alpha} t) \nonumber\\
\end{eqnarray}

Now lets evaluate these for each of our models of interest, working close to the solvable points.

\subsection{Ising model}
At the exactly solvable point, $\langle M^z(t) M^z(0)\rangle$ will only have a Fourier component at zero frequency, because $Z$ operators will not produce any transitions between states. 
However $X$ and $Y$ operators will produce spin flips. $|\alpha \rangle$ will thus be a state with a single flipped spin. The energy cost at the exactly solvable point will be $z$, where $z$ is the coordination number of the lattice, producing a sharp $\delta$ function peak at frequency $\omega = z$. Away from the exactly solvable point the spins will acquire dispersion $E(\vec{k})$. We assume that because of the large (effectively infinite) speed of light, while the light can supply energy it cannot supply momentum, so the transitions involved can only be to states with zero total momentum. For the Ising model $|\alpha\rangle$ has to be the $k=0$ spin flip state, in which case we will observe a {\it sharp} feature in linear response at the corresponding frequency. This sharp feature is indicative of the existence of a sharp excitation at zero momentum. This assumes that the energy of a single $k=0$ spin flip does not overlap the multi-excitation continuum, so there is no decay channel available for the single spin flip state. This assumption should be safe as long as we are away from the critical point. 

\subsection{Two dimensional toric code}
Lets start by examining the $M^z$ correlator. A $Z$ operator will anticommute with two $A$ operators. Thus, $|\alpha \rangle$ will be a state with two `spinons.' At the exactly solvable point it will have energy $2$ and a sharp signal in the spectrum at frequency $2$. If we perturb about the exactly solvable point the spinons will acquire dispersion $E(\vec{k})$. Let us assume as before that the light does not supply appreciable momentum, so $|\alpha\rangle$ has to be a state with zero total momentum. Nonetheless the two spinons can have momentum $\pm k$ and so the state can have energy $2E(k)$ \footnote{For large enough $k$ the energy of the `two spinon' state may overlap the multi-excitation continuum, in which case new decay channels may open up, but this does not alter our conclusions.}. So away from the exactly solvable point here will be no sharp signature in linear response, but only a broad continuum. The behavior of the $ M^x$ correlator is analogous with the only difference an altered dispersion relation. Finally, a $Y$ operator will create two $A$ spinons and two $B$ spinons.  At the solvable point there will be a sharp signal at frequency $2+2J$. Away from the solvable point it will blur into a continuum. 

The observation of a broad continuum in linear response (with any polarization) indicates that $Z$ polarized light does not create a single sharp excitation at zero momentum. However, it cannot distinguish between the case where it creates a {\it pair} of sharp quasiparticles (as is the case here, at least in the frequency range where the two spinon state does not overlap the multi-spinon continuum), and the case where there is no sharp excitation because e.g. there are no good quasiparticles. We will see in Section \ref{sec: results} how 2DCS can disambiguate between these possibilities. 

\subsection{X-cube}
A $Z$ operator will anticommute with the four $B$ operators containing that link. Let us suppose it is an $\hat x$ directed link for specificity. It will thus create a multiplet of four cube excitations (fractons). At the exactly solvable point this will produce a sharp signal at frequency $4K$. Now lets consider the generic case, where we perturb about the solvable point. As before we restrict to states $|\alpha\rangle$ with total momentum zero. Now pairs of two fractons are planeons (mobile excitations restricted to move in  either the $xy$ plane or the $xz$ plane, depending on how we group the fractons). These will acquire dispersion upon turning on perturbations so there will be a range of possible energies for $|\alpha\rangle$. As before there will be nothing sharp in linear response. 

Now lets consider the $M^x$ correlator. An $X$ type operator will anticommute with four $A$ type operators (two for each vertex sharing that link). The excitation at each vertex is a lineon - i.e. when we turn on perturbations each lineon can move freely, but only in the direction of the link. Nevertheless, the lineons will acquire dispersion. Naively we might think that this means there will  not be any sharp signal in linear response, but this is too fast - actually there is a sharp signal here. Namely, the lineons, being one dimensional particles, will have a one dimensional Van Hove singularity in their density of states, which will diverge near the gap edge as $1/\sqrt{E}$. If we assume that the energy of the lineons is minimized at $k=0$ (which is a highly plausible assumption) then this should be readily observable in linear response spectroscopy, and should provide a clean signature of the existence of {\it one dimensional} excitations in this (otherwise three dimensional) system. 

Finally, a $Y$ operator will anticommute with four $B$ operators and four $A$ operators, and will thus excite four `fractons' (which could be grouped into two planeons), as well as four vertex operators (grouped into two lineons). Again the lineons will have a crisp signature through their one dimensional density of states, and so the $M^y$ correlator will also have clear signatures of the existence of one dimensional particles.

In all the above, we have implicitly worked in the infinitesimal neighborhood of the exactly solvable point where e.g. an $X$ operator only creates a single pair of lineons. Away from the solvable point there will also be some admixture of four lineon creation and higher multiparticle processes. (One way to see this is to perform a Schrieffer Wolff transformation \cite{SchriefferWolff} to remove the perturbation, and to note that the new `A' operators have some admixture of highly multiparticle operators in the original basis). These `multiparticle absorbtion' channels should be added on to the absorbtion spectrum discussed above, and will give rise to additional multiparticle continua. However, the existence of multiparticle continua will not alter the fact that there is a one dimensional Van Hove singularity, which should be clearly detectable in absorbtion experiments. (In addition, the threshold frequency for the multiparticle continua will generically be higher than the threshold frequency for the two lineon continuum, so the above analysis will be strictly accurate in the frequency range where the light supplies enough energy to create two lineons, but not to create more than two excitations).  

Thus, type I gapped fracton phases which support one dimensional (i.e. lineon) excitations will have a clear signature in linear response, in the form of a one dimensional Van Hove singularity in the absorption spectrum of an otherwise three dimensional material. The only challenge (for crisply identifying this signature as coming from lineons) is to show that the signature does \textit{not} come because e.g. the material has the structure of a system of weakly coupled one dimensional wires. This should be straightforward to do \textit{in combination} with some alternative experiment e.g. transport. In transport experiments (performed at non-zero temperature, so that there is transport) a system of coupled wires would have a clear preferred axis, whereas a fracton system with lineons would not, since there are lineons that can move along any lattice axis. An alternative (purely spectroscopic) way to tell is that there will not be any signature of `one dimensional' physics in the $M^z$ correlator.

\subsection{Haah code}
Now $X$ and $Z$ correlators behave the same way. Acting with either $X$ or $Z$ creates four excitations in a `tetrahedral' arrangement (four fractons), with energy cost $4$ (or $4\Lambda$) at the exactly solvable point. These cannot be grouped into mobile excitations (away from the solvable point), so we continue to have a sharp `delta function' signature at frequency $4$ even away from the exactly solvable point. This remains the case even if we e.g. turn on a magnetic field and sweep the angle. 

The existence of a sharp robust signature in linear response is tempting to identify as diagnostic of a type II fracton phase. However we should bear in mind that something similar could show up if we had e.g. well isolated two level systems in the problem, or indeed if local fields created sharp individual excitations (as in the Ising model) which are then constrained to have zero momentum. As such, the observation of a sharp robust signal in linear response alone is not sufficient to conclude that one is dealing with a type II fracton phase. 

\subsection{Summary of linear response}
To conclude: type I fracton phases (X-cube) do have a crisp diagnostic in linear response, in the form of a one dimensional Van Hove singularity in the absorbtion spectrum which is present only for certain polarizations of the incident light. This is a signature that the phase contains fractionalized `lineon' excitations i.e. excitations that can only move in one dimension. In contrast, neither conventional gapped spin liquids (toric code) nor type II fracton phases (Haah code) have crisp signatures in linear response. To diagnose these phases we will need to turn to 2DCS and non-linear response. 

\section{2DCS and nonlinear response}
\label{sec: nlresponse}
We consider a 2DCS experiment where two $\delta$ function pulses of magnetic fields are applied at time $0$ and $\tau$.
Two magnetic fields are linearly polarized along $\alpha$ and $\beta$ direction, respectively:
\begin{equation}
\mathbf{B}(s) = \hat{e}_\alpha B_\alpha  \delta(s) + \hat{e}_\beta B_\beta  \delta(s-\tau).
\end{equation}
The pulse induced magnetization $M_\gamma(t+\tau)$ is measured at later time $t+\tau$:
\begin{widetext}
\begin{align}
&M_\gamma(t+\tau) / N_{\mathrm{spin}} = \chi_{\gamma \alpha} (t+\tau) B_{\alpha} +  \chi_{\gamma \beta} (t) B_{\beta} + \chi_{\gamma \alpha \alpha}(t+\tau, t+\tau) B_{\alpha} B_{\alpha} + \chi_{\gamma \beta \beta}(t, t) B_{\beta} B_{\beta} + \chi_{\gamma \alpha \beta}(t, t+\tau)  B_{\alpha} B_{\beta} \nonumber\\
&+ \chi_{\gamma \alpha \alpha \alpha}(t+\tau) B_{\alpha} B_{\alpha} B_{\alpha} + \chi^{3}_{\gamma \beta \beta \beta}(t) B_{\beta} B_{\beta} B_{\beta} + \chi_{\gamma \alpha \alpha \beta}(t, t+\tau, t+\tau) B_{\alpha} B_{\alpha} B_{\beta} + \chi_{\gamma \alpha \beta \beta}(t, t, t+\tau) B_{\alpha} B_{\beta} B_{\beta} + \mathcal{O}(B^4),
\end{align}
where $N_{\mathrm{spin}}$ is the total number of spins on a lattice.
There is no sum over repeated indices.
The canonical 2DCS experiment extracts the \emph{nonlinear} response by subtracting off the signal observed in the presence of either pulse alone. This leaves us with 
\begin{align}
M^{\mathrm{nonlinear}}_{\gamma}(t+\tau)/N_{\mathrm{spin}} = \chi_{\gamma \alpha \beta}(t, t+\tau)B_{\alpha} B_{\beta} +  \chi_{\gamma \alpha \alpha \beta}(t, t+\tau, t+\tau) B_{\alpha} B_{\alpha} B_{\beta} + \chi_{\gamma \alpha \beta \beta}(t, t, t+\tau) B_{\alpha} B_{\beta} B_{\beta} + \mathcal{O}(B^4)
\label{eq:chidef}
\end{align}
\end{widetext}

Now let us consider two possible experiments.
Experiment $\mathrm{I}$ has $(\alpha, \beta, \gamma) = (x, x, x)$.
Experiment $\mathrm{II}$ has $(\alpha, \beta, \gamma) = (z, z, z)$.
Other combinations of polarizations are of course possible, and may be interesting to consider, but these two are sufficient to provide an unambiguous diagnostic of spin liquids, both conventional and fractonic.
If we consider the models on a square and cubic lattice, the second order susceptibility $\chi_{\gamma\alpha\beta}(t,t+\tau) = 0$ for those two experiments because we need even number of $X$ or $Z$ operators to pair up creation and annihilation of excitations.
Then, the leading contributions to the nonlinear signals are the third order susceptibilities, $\chi_{\gamma \alpha \alpha \beta}(t,t+\tau,t+\tau)$ and $\chi_{\gamma \alpha \beta \beta}(t,t,t+\tau)$.
They are related to four-point correlation functions via the generalized Kubo formula \cite{choileekim}:
\begin{widetext}
\begin{align}
\chi_{\gamma \alpha \alpha \beta}(t,t+\tau,t+\tau) 
&= - \frac{i}{N_{\mathrm{spin}}} \Theta(t)\Theta(\tau)
\langle [[[M_{\gamma}(t+\tau),M_{\beta}(\tau)],M_{\alpha}(0)],M_{\alpha}(0)]\rangle, \\
\chi_{\gamma \alpha \beta \beta}(t,t,t+\tau)
&= - \frac{i}{N_{\mathrm{spin}}} \Theta(t)\Theta(\tau)
\langle [[[M_{\gamma}(t+\tau),M_{\beta}(\tau)],M_{\beta}(\tau)],M_{\alpha}(0)]\rangle,
\end{align}
where $\Theta$ is Heaviside step function (equal to one when the argument is larger than zero), and the operators have been time evolved (Heisenberg picture) with respect to the system Hamiltonian.
Note that $M_{\beta}$ is always applied at time $\tau$, $M_{\alpha}$ at time $0$, and $M_{\gamma}$ at time $t+\tau$. 

Let us define 
\begin{eqnarray}
R_{abcd} &=& \langle M_a(t_a) M_b (t_b) M_c(t_c) M_d (t_d) \rangle \nonumber\\ &=& \sum_{\mu \nu \lambda} \langle 0 | M_a | \mu \rangle \langle \mu | M_b | \nu \rangle \langle \nu | M_c | \lambda \rangle \langle \lambda | M_d | 0 \rangle \exp\left( \frac{i}{\hbar} (E_{\mu} (t_b-t_a) + E_{\nu} (t_c - t_b) + E_{\lambda} (t_d-t_c)\right),
\label{eq:R}
\end{eqnarray}
where in the last line we have reverted to the Schrodinger picture, and $E_{\mu,\nu,\lambda}$ are the energy difference between the excited state and the ground state.
Now we can write 
\begin{align}
&\chi_{\gamma \alpha \alpha \beta}(t,t+\tau,t+\tau)  = - \frac{i}{N_{\mathrm{spin}}} \Theta(t)\Theta(\tau) \left[R_{\gamma \beta \alpha \alpha} - R_{\beta \gamma \alpha \alpha} - 2R_{\alpha \gamma \beta \alpha} + 2 R_{\alpha \beta \gamma \alpha} + R_{ \alpha \alpha \gamma \beta} - R_{ \alpha \alpha \beta \gamma} \right],\\
&\chi_{\gamma \alpha \beta \beta}(t,t,t+\tau) = - \frac{i}{N_{\mathrm{spin}}} \Theta(t)\Theta(\tau) \left[R_{\gamma \beta \beta \alpha} - 2 R_{\beta \gamma \beta \alpha} + R_{\beta \beta \gamma \alpha} - R_{\alpha \gamma \beta \beta} + 2 R_{\alpha \beta \gamma \beta} - R_{\alpha \beta \beta \gamma } \right].
\end{align}
We always have $t_{\gamma} = t+\tau$, $t_{\beta} = \tau$ and $t_{\alpha} = 0$. Let $S_{abcd}$ be just the matrix element part of $R_{abcd}$ (i.e. without the phases in Eq.~(\ref{eq:R})).
Then we can write
\begin{align}
\chi_{\gamma \alpha \alpha \beta}(t,t+\tau,t+\tau)  &= - \frac{i}{N_{\mathrm{spin}}} \Theta(t)\Theta(\tau) \sum_{\mu\nu\lambda} \left[S_{\gamma \beta \alpha \alpha} e^{-i(E_{\mu} t + E_{\nu} \tau) } - S_{\beta \gamma \alpha \alpha} e^{i((E_{\mu}-E_{\nu}) t - E_{\nu} \tau) }- 2S_{\alpha \gamma \beta \alpha}e^{i((E_{\mu}-E_{\nu}) t + (E_{\mu}-E_{\lambda}) \tau) }\right. \nonumber\\
&\left.+ 2 S_{\alpha \beta \gamma \alpha} e^{i((E_{\nu}-E_{\lambda}) t + (E_{\mu}-E_{\lambda}) \tau) }+ S_{ \alpha \alpha \gamma \beta}e^{i( (E_{\nu} - E_{\lambda})t + E_{\nu} \tau)} - S_{ \alpha \alpha \beta \gamma} e^{i( E_{\lambda}t + E_{\nu} \tau)}  \right],\label{chi31}\\
\chi_{\gamma \alpha \beta \beta}(t,t,t+\tau) &= - \frac{i}{N_{\mathrm{spin}}} \Theta(t)\Theta(\tau)\sum_{\mu\nu\lambda} \left[S_{\gamma \beta \beta \alpha} e^{-i(E_{\mu} t + E_{\lambda} \tau)}- 2 S_{\beta \gamma \beta \alpha}e^{i((E_{\mu} -E_{\nu})t - E_{\lambda} \tau)} + S_{\beta \beta \gamma \alpha} e^{i((E_{\nu} -E_{\lambda})t - E_{\lambda} \tau)}\right.\nonumber\\
&\left.- S_{\alpha \gamma \beta \beta} e^{i((E_{\mu} -E_{\nu})t + E_{\mu} \tau)} + 2 S_{\alpha \beta \gamma \beta} e^{i((E_{\nu} -E_{\lambda})t + E_{\mu} \tau)}- S_{\alpha \beta \beta \gamma} e^{i(E_{\lambda} t + E_{\mu} \tau)} \right], \label{chi32}
\end{align}
where we have set $\hbar = 1$, and the expressions are now in the Schrodinger picture.
Here $S_{abcd}$ is the implicit function of $\mu$, $\nu$, and $\lambda$.
We note that the experimentally observed signal will be $\chi_{\gamma \alpha \alpha \beta} (t,t+\tau,t+\tau) B_{\alpha}^2 B_{\beta}  + \chi_{\gamma \alpha \beta \beta} (t,t,t+\tau) B_{\alpha} B_{\beta}^2$.
Thus both third-order susceptibilities could, in principle, be extracted by individually varying the strength of the two applied pulses. We will thus generally present the two third-order susceptibilities separately, even though what will be observed will of course be the sum of the two. 

There is one subtlety to note here. In what follows we will generally be looking at the response in frequency space, and the Fourier transform of the terms in square brackets is the quantity of interest. However, because the susceptibility is multiplied by step functions, the observable signal will consist of the `interesting' signal (the terms in square brackets) convolved with $(\delta(\omega_t) +  \mathcal{P} i/\omega_{t})(\delta(\omega_{\tau}) + \mathcal{P} i/\omega_{\tau}) = \delta(\omega_t) \delta(\omega_{\tau}) - \mathcal{P} \frac{1}{\omega_t \omega_{\tau}} + i (\delta(\omega_t) \mathcal{P}\frac{1}{\omega_{\tau}} +\delta(\omega_{\tau}) \mathcal{P}\frac{1}{\omega_{t}}) $, where $\mathcal{P}$ denotes principal value. How this effects the observable signal will be noted below, where appropriate. 
\end{widetext}

\section{Results}
\label{sec: results}

We now consider each of our models, in each of the two potential polarizations, $(\alpha,\beta,\gamma) = (x,x,x)$ and $(\alpha,\beta,\gamma) = (z,z,z)$.
\subsection{Ising model}
\begin{figure*}[t]
\centering
\includegraphics[width=\linewidth]{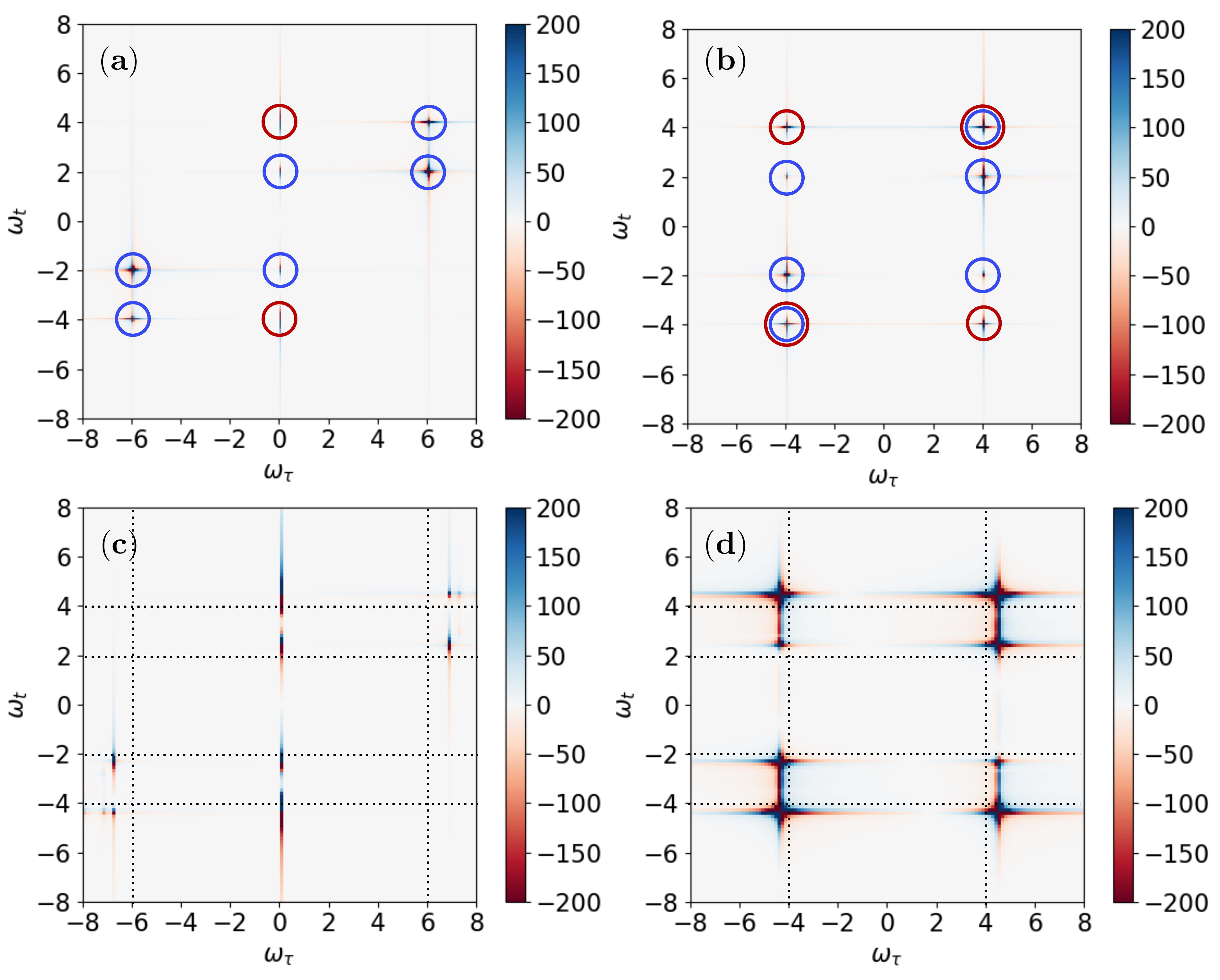}
\caption{Imaginary part of 2D Fourier transform $\mathcal{F}_{t,\tau}$ of the third-order susceptibilities $\chi_{xxxx}$ for 2D Ising model (a, b) without perturbations and (c, d) with generic perturbations.
(a) $\mathrm{Im}\,\mathcal{F}_{t,\tau}[\chi_{xxxx}(t,t+\tau,t+\tau)]$ and (b) $\mathrm{Im}\,\mathcal{F}_{t,\tau}[\chi_{xxxx}(t,t,t+\tau)]$
of 2D Ising model show sharp pointlike signals due to the bound state of two magnons (marked with blue circles) and two separated magnons (marked with red circles).
(c) $\mathrm{Im}\,\mathcal{F}_{t,\tau}[\chi_{xxxx}(t,t+\tau,t+\tau)]$ and (d) $\mathrm{Im}\,\mathcal{F}_{t,\tau}[\chi_{xxxx}(t,t,t+\tau)]$ of the perturbed Ising model exhibit finite shift of the location of the signals by the half-bandwidth $2w = 0.4$. The dotted lines guide the location of unperturbed signals.
The perturbations also broaden the red circled signals coming from two magnons with nonzero momenta $\pm\mathbf{k}$.}
\label{fig: Ising2dcs}
\end{figure*}

We start by working in two spatial dimensions. In this case, at leading order close to the solvable point, the experiment $\mathrm{II}$ with $zzz$ polarization does not produce any transitions or lead to any nonlinear signal. So we can focus on the experiment $\mathrm{I}$ with $xxx$ polarization. 
There is only one matrix element to be evaluated. The states $|\mu\rangle$ and $|\lambda\rangle$ both contain a single flipped spin with zero momentum, with energy $E_{\mu} = E_{\lambda} = 4 + \delta_0$, where $\delta_0$ denotes corrections to the energy coming from perturbations \footnote{In principle away from the solvable point a single $X$ operator can also produce multiple excitations, which will yield additional additive contributions to the non-linear response, and will also open up intrinsic relaxation channels producing line broadening. An incorporation of such effects would be an interesting problem for the future but is beyond the scope of the present work. In general those multi-excitation contributions will occur at a higher frequency and our discussion should be strictly accurate for energies below this multi-excitation continuum}.
Meanwhile $|\nu\rangle$ contains either zero or two flipped spins which may be adjacent but do not need to be. Thus we have $E_{\nu} = 0$, $E_{\nu} = 8 + \delta_k$ or $E_{\nu} = 6 + \delta'$, where the correction $\delta_k$ is the kinetic energy contributions coming from configurations where the two flipped spins have momentum $\pm \vec{k}$, and $\delta'-2 \delta_0$ is the `binding energy' for two adjacent spin flips (with zero momentum). In the former case we need to sum over $\vec{k}$. Note that because $E_{\nu}$ is dispersive, the signal will be broadened in the frequency directions corresponding to $E_{\nu}$. However it will be sharp in all other directions, modulo the broadening from convolution discussed above. We further note that the pathway with $E_{\nu}=0$ just corresponds to linear response done twice, and cannot contribute to the nonlinear response (See Appendix for detailed justification, but one can also simply note that the contribution coming from such a pathway would scale as the system size, whereas the nonlinear susceptibility must of course be independent of the system size.). We can therefore focus on the terms with $E_{\nu} = 8 + \delta_k$ or $E_{\nu} = 6 + \delta'$. 

Let $(\omega_t, \omega_\tau)$ be the frequencies conjugate to $t$ and $\tau$, respectively. Then the channel with $E_{\nu} = 6 + \delta'$ will give rise to sharp (delta function) signals at frequencies $\omega_{t, \tau}$ equal to $2+\delta'-\delta_0$, $4 + \delta_0$, or $6 + \delta' + \delta_0$. Thus, signals will appear with a spacing in frequency space equal to roughly {\it half} the linear response gap. It is important to bear in mind that this occurs in a clearly non-spin-liquid system, so the appearance of signals at a fraction of the linear response gap is not in itself a signal of fractionalization. 


There will also be contributions from intermediate states with $E_{\nu} = 8+\delta_k$, These will be broadened along one direction (that corresponding to $E_{\nu}$ but will be sharp in the other direction. This will give rise to streaks in the $\omega_t$ direction, with and without offset $4$ in the $\omega_{\tau}$ direction, to a streak in the $\omega_{\tau}$ direction with offset $4$ in the $\omega_t$ direction, and also to a diagonal streak with offset $4$ in the $\omega_t$ direction - see Fig. \ref{fig: Ising2dcs} for an illustration. This is our `control' example of a non-spin liquid. Features that appear herein cannot be viewed as spin liquid signatures. 

Meanwhile in three dimensions, the possible energies are shifted to $E_{\mu} = 6 = E_{\lambda}$ and $E_{\nu} = \{0,10 + \delta', 12+\delta_k\}$ and the signatures are analogous to the two dimensional case.  

The actual results in Fig.\ref{fig: Ising2dcs} are produced via a direct calculation from Eq.\ref{chi31}, \ref{chi32} in two dimensions, including matrix elements. Subfigures (a, b) corresponds to the exactly solvable point $\lambda=0$. In subfigures (c, d) we have assumed that the excited states are plane waves, and have calculated matrix elements accordingly. Excited states containing two flipped spins are a symmetrized product of two plane waves, as appropriate for bosonic excitations. For a detailed discussion of the calculation, see the Appendices.

We now have to discuss an annoying subtlety associated with the fact that the experimentally observed signal gets convolved with $\delta(\omega_t) \delta(\omega_{\tau}) - \mathcal{P} \frac{1}{\omega_t \omega_{\tau}} + i (\delta(\omega_t) \mathcal{P}\frac{1}{\omega_{\tau}} +\delta(\omega_{\tau}) \mathcal{P}\frac{1}{\omega_{t}})$, because the Fourier transforms only run over the positive time axes. This produces a weak $1/\omega$ broadening in both directions (from convolution with the second term), and a somewhat stronger $1/\omega$ broadening along both axes (from convolution with the third and fourth terms). This `broadening from convolution' can make it hard to see the more physical broadening from dispersion. Within our present approximations, the physical part of the spectrum is pure imaginary in Fourier space (see appendices), and given that the strongest part of the broadening from convolution comes with an extra factor of $i$, it may simply be removed by taking the imaginary part of the non-linear susceptibility - this is done in Fig. \ref{fig: Ising2dcs}, and will be done throughout for all the models we study. However, once decay of the excitations is re-introduced, whether through coupling to extraneous degrees of freedom such as phonons, or through decay into multi-excitation sectors (at frequencies overlapping the multi-magnon continuum), then the `physical' part of the signal will in general become complex, and the `broadening from convolution' problem will become unavoidable. However, the $1/\omega$ broadening from convolution still leaves the intensity of the signal sharply peaked where it would have been, so if one simply applies a high-pass filter on intensity then this may suffice to deal with the problem. 

 \subsection{Toric code} \label{ssec:TC}
 For the toric code (our paradigmatic example of a conventional gapped spin liquid), both polarization configurations yield interesting results. 
A Z operator applied to any link creates a pair of vertex excitations residing on the two vertices adjacent to that link. (More generally, the end points of strings of Z operators produce vertex excitations). Meanwhile, X operators applied to a link create a pair of plaquette excitations on the two plaquettes containing that link. More generally, strings of X operators create plaquette excitations are the ends. Both vertex and plaquette excitations are mobile but are not locally creatable (the locally creatable things are pairs of vertex or plaquette excitations). We will restrict the analysis to the sector where each $Z$ or $X$ operator locally creates only two plaquette or vertex excitations and will ignore mixing with the `many excitation' sector. `Multiexcitation' channels will make additional contributions to the signal, but as we will see unambiguous diagnostics appear already at leading order (and will be there regardless of what additional signatures appear from multiexcitation pathways). As usual, neglect of multi-excitation pathways should also be safe at frequencies below the multi-excitation gap.  We will allow all excitations to have dispersion (i.e. we will implicitly perturb about the solvable point). 

\begin{figure*}[t]
\centering
\includegraphics[width=\linewidth]{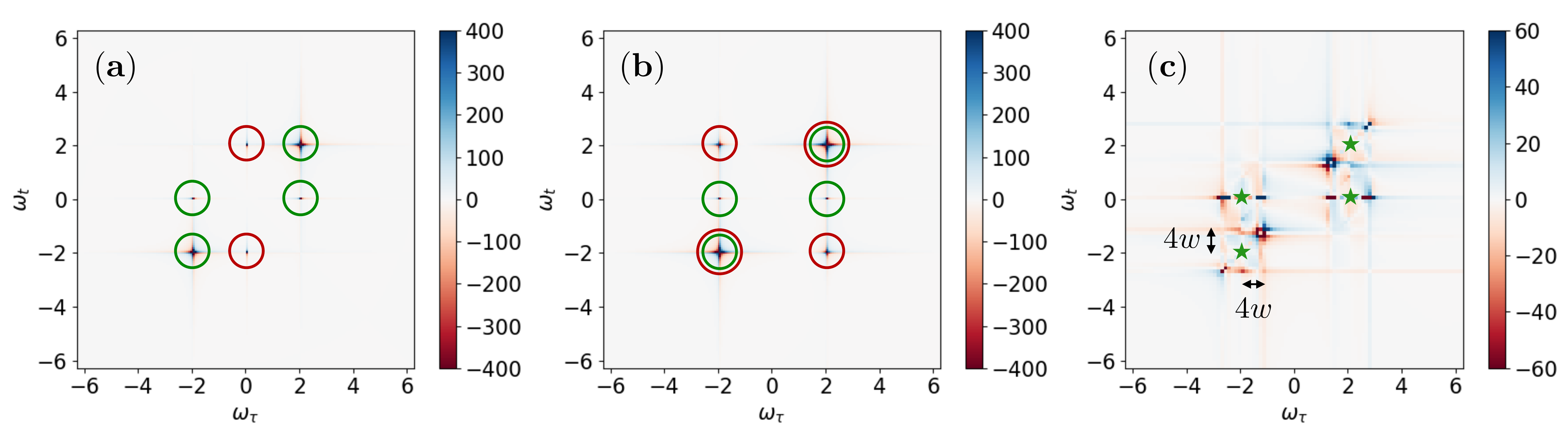}
\caption{Imaginary part of 2D Fourier transform $\mathcal{F}_{t,\tau}$ of the third-order susceptibilities $\chi_{xxxx}$ for 2D toric code (a, b) without perturbations and (c) with generic perturbations.
(a) $\mathrm{Im}\,\mathcal{F}_{t,\tau}[\chi_{xxxx}(t,t+\tau,t+\tau)]$ and (b) $\mathrm{Im}\,\mathcal{F}_{t,\tau}[\chi_{xxxx}(t,t,t+\tau)]$
of the toric code show sharp pointlike signals due to plain motion of a pair of two plaquettes (marked with green circles, $E_\mu=E_\nu=E_\lambda$ process in Sec.~\ref{ssec:TC}) and quantum processes involving two pairs of plaquettes (marked with red circles, $E_\nu = E_\mu + E_\lambda$ process in Sec.~\ref{ssec:TC}).
The signals at $(\omega_\tau, \omega_t) = (\pm 2, 0)$ are the distinct signatures of the model having deconfined excitations because the signals suggest vanishing string tension between two plaquettes of the toric code.
(c) $\mathrm{Im}\,\mathcal{F}_{t,\tau}[\chi_{xxxx}(t,t+\tau,t+\tau)]$ of the perturbed toric code.
$\mathrm{Im}\,\mathcal{F}_{t,\tau}[\chi_{xxxx}(t,t,t+\tau)]$ shows indistinguishably equal nonlinear spectrum.
The nonlinear signals due to the process involving only two plaquettes ($E_\mu=E_\nu=E_\lambda$) are shown.
Unlike Ising model, presence of the dispersive deconfined excitations yield clear spread of the signals by the full bandwidth of the single plaquette dispersion $4w = 0.8$. Green stars mark locations of the signals without the perturbations.}
\label{fig:TC}
\end{figure*}

 \subsubsection{$(\alpha, \beta, \gamma) = (x, x, x)$}
Here $|\lambda\rangle$ and $|\mu\rangle$ contain a pair of plaquette excitations with net momentum zero. We have $E_{\mu} = 2J + 2 \delta_k$ and $E_{\lambda} = 2J+2\delta_{k'}$, where the plaquette excitations have momenta $\pm k$ and $\pm k'$ respectively, and a corresponding kinetic energy $\delta_k (\delta_{k'})$.  Meanwhile, $|\nu\rangle$ could contain any of: zero plaquette excitations ($E_{\nu} = 0$), two plaquette excitations ($E_{\nu} = E_{\mu} = E_{\lambda})$, or four plaquette excitations ($E_{\nu} =  E_{\mu} + E_{\lambda}$). All momenta have to be summed over. As before, the channel with $E_{\nu} = 0$ corresponds to linear response done twice and cannot make a (properly extensive) contribution to the non-linear response, and will therefore be ignored. (In the appendix we explicitly show how this channel cancels to give zero contribution to the non-linear response). We will therefore focus on the channels with $E_{\nu} = E_{\mu} = E_{\lambda}$, or $E_{\nu} =  E_{\mu} + E_{\lambda}$. We are interested only in signals that are sharp in at least one direction, after allowing dispersion. 


 We begin by considering the sequence with $E_{\nu} = E_{\mu} = E_{\lambda}$. This channel corresponds to creation of a string with visons (plaquette excitations) at the two ends, followed by elongation of the string, followed by annihilation of the string, with the fact that all states have the same energy being a consequence of the string having zero line tension. Since zero line tension for strings is equivalent to deconfinement, we can reasonably expect signals from this channel to contain signatures of deconfinement. Now the first and last terms in Eq.(\ref{chi31}, \ref{chi32}) give rise to diagonal stripes (outside the two vison gap), whereas the second and fifth terms in Eq.\ref{chi31}), together with the second, third, fourth and fifth terms in Eq.\ref{chi32} collectively give rise to a sharp signal along the $\omega_{\tau}$ axis in the two dimensional Fourier transform (again, above the two vison gap). Both these features (the sharp streak along the $\omega_{\tau}$ axis, and the streak along the diagonals {\it with no offset} are signatures of deconfinement, at least in conjunction with the lack of a sharp signal in linear response. The crucial aspect here is that the streaks {\it have no offset} - while the signal is only present outside the two vison gap, the extrapolation of the streak goes through the origin. This is a consequence of the fact that, as mentioned, a `string' hosting visons at the two ends can change length with no change in energy - a signature of deconfinement. Another way to view it is that each $X$ operator creates a pair of visons, but the visons are their own antiparticles so the application of an $X$ operator to a two vison state can leave us in a two vison state. That is, the first pulse creates a pair of visons with momentum $\pm k$, the second creates another pair of visons with momentum $\pm k$ (one of which annihilates one of the visons in the original pair), and the third pulse then returns us to the ground state. Again, the fact that a local operator creates a pair of sharply defined $Z_2$ charged excitations is a signature of fractionalization. In contrast, in the two dimensional Ising model the equivalent streaks had an offset, indicating that the second $X$ operator necessarily changed the energy of the state. In the one dimensional Ising model, equivalent signals do arise \cite{ArmitageWan} - but then the one dimensional Ising model also has deconfined fractionalized excitations (the domain walls).

 In Fig. \ref{fig:TC}, we present computations of the signal including appropriate matrix elements in \ref{chi31}, \ref{chi32}, both at and away from the solvable point. The wavefunctions are taken in real space at the exactly solvable point, and are assumed to be symmetrized products of plane waves away from the solvable point. Full details of the calculation are presented in the appendices. As we see, the features present in the explicit calculation are precisely those expected by inspection of \ref{chi31}, \ref{chi32}, and knowledge of the toric code.

Now there is the channel with $E_{\nu} = E_{\mu} + E_{\lambda}$, where $E_{\mu}$ and $E_{\lambda}$ may be individually varied. Naively, this produces a contribution that scales like the square of the volume, so to leading order the contributions from this channel must cancel. However, this may leave a non-trivial extensive piece behind - after all this channel is not just linear response done twice. Explicit calculation of this channel is impractically tedious away from the exactly solvable point - the states $|\mu\rangle$ and $|\lambda\rangle$ are a symmetrized product of two plane waves, whereas the state $|\nu\rangle$ is a symmetrized product of four plane waves ($4!$ terms in the sum), and altogether there are $2*4!*2 = 96$ possible terms to evaluate and sum. However, there is also no good reason to believe this channel should have clear diagnostics of deconfinement, and we have already found clear signatures of deconfinement from the channel with $E_{\mu} = E_{\nu} = E_{\lambda}$. The only concern might be that the channel with $E_{\nu} = E_{\nu} + E_{\lambda}$ might somehow cancel the clear signatures we found coming from the channel with $E_{\mu} = E_{\nu} = E_{\lambda}$. We have verified that this does not happen at the exactly solvable point by explicit calculation (see Appendix, and also Fig.\ref{fig:TC}(a)). Accordingly, we believe it is safe to ignore the contributions from this channel - there may be additional contributions to the non-linear response from it, but they do not cancel the contributions from the $E_{\mu} = E_{\nu} = E_{\lambda}$ channel, and those already contain clear signatures of deconfinement. 


Thus, the \textit{combination} of linear response and 2DCS data can provide clean fingerprints of deconfined fractionalized excitations. If linear response does not show any sharp features this could be because local fields create multiplets of dispersive fractionalized excitations (as in the toric code), or it could just mean that they create excitations that are not good quasiparticles. However, if 2DCS also shows a sharp stripe along the $\omega_{\tau}$ axis and along the diagonal, then this indicates that local fields in fact create a pair of quasiparticles which are their own antiparticles, such that a second pulse can leave us in the `two quasiparticle' sector. This is the case for deconfined fractionalized excitations that live at the ends of strings with zero line tension - the second pulse just elongates the string - but it is not the case if there are non-fractionalized excitations which are just not good quasiparticles. 

The analysis of the experiment for fields in the $z$ direction proceeds analogously - fields create vertex excitations instead of plaquette excitations, but we obtain identical results. 

Thus, from linear response, we can learn whether local fields can create sharply defined individual quasiparticles. If the answer is no, then using 2DCS we can determine if local fields actually create pairs of deconfined fractionalized quasiparticles.




 \subsection{X-cube} \label{ssec:Xcube}
 We now consider 2DCS on the X-cube phase. Now an $X$ operator applied to a link creates a pair of vertex excitations (lineons) on the two vertices adjacent to that link. These lineons can move freely, but only in one dimension (the dimension of the link). 
 
 A $Z$ operator applied to a link creates a quartet of cube excitations (fractons) on the four cubes sharing that link. These fractons are individually immobile, but can move in pairs as planeons (i.e. can move freely in the plane orthogonal to the long axis of the planeon).
 
 \subsubsection{$(\alpha, \beta, \gamma) = (x,x,x)$} \label{sssec:lineon}
 
\begin{figure*}[t]
\centering
\includegraphics[width=\linewidth]{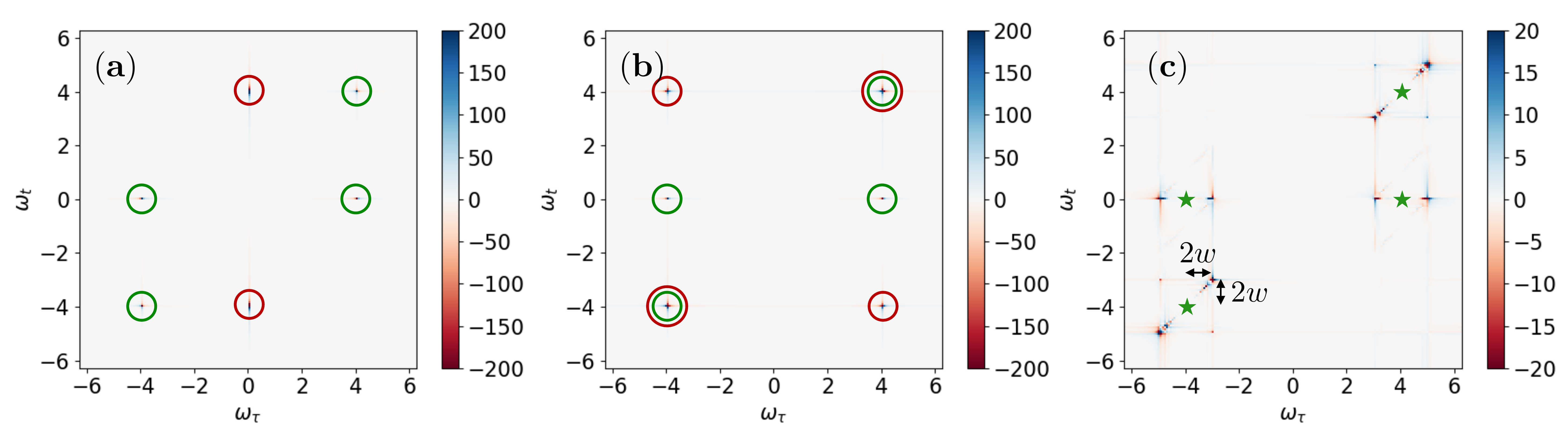}
\caption{Imaginary part of 2D Fourier transform $\mathcal{F}_{t,\tau}$ of the third-order susceptibilities $\chi_{xxxx}$ for X-cube model (a, b) without perturbations and (c) with generic perturbations.
Qualitatively, the Fourier-transformed $\chi_{xxxx}$ of X-cube model is similar to those of the toric code.
(a) $\mathrm{Im}\,\mathcal{F}_{t,\tau}[\chi_{xxxx}(t,t+\tau,t+\tau)]$ and (b) $\mathrm{Im}\,\mathcal{F}_{t,\tau}[\chi_{xxxx}(t,t,t+\tau)]$
of X-cube model show sharp pointlike signals suggesting the deconfinement, \textit{i.e.}, a pair of two lineons move with zero string tension (marked with green circles, $E_\mu=E_\nu=E_\lambda$ process in Sec.~\ref{sssec:lineon}). Red circled signals are due to quantum processes involving four lineons ($E_\nu = E_\mu + E_\lambda$ process in Sec.~\ref{sssec:lineon}).
(c) $\mathrm{Im}\,\mathcal{F}_{t,\tau}[\chi_{xxxx}(t,t+\tau,t+\tau)]$ of the perturbed X-cube model.
$\mathrm{Im}\,\mathcal{F}_{t,\tau}[\chi_{xxxx}(t,t,t+\tau)]$ is almost the same.
Only the signals relevant to $E_\mu=E_\nu=E_\lambda$ process are shown.
Similar to the toric code, dispersive lineons exhibit clear spread of the signals by the full bandwidth of a lineon dispersion $2w = 1$. Green stars mark locations of the peaks without the perturbations.}
\label{fig:lineon}
\end{figure*}
 
 Now $|\mu\rangle$ and $|\lambda\rangle$ both contain a single pair of lineons with zero total momentum (and with associated Van Hove singularity in the DOS). Meanwhile, $|\nu\rangle$ can contain either zero lineons, a single pair of lineons (in which case $E_{\mu} = E_{\nu} = E_{\lambda}$, or two pairs of lineons. The lineons are deconfined fractionalized excitations which are their own antiparticles, much like the visons in the toric code. The analysis parallels the analysis of the toric code, with the one exception that the DOS is that of a one dimensional system. In particular, we will have the same diagnostics of deconfinement as in the toric code. These include streaks along the $\omega_{\tau}$ axis and along the diagonal, with no offset. This is illustrated in Fig.\ref{fig:lineon}. Full calculations are in the appendices - we have performed exact calculations at the solvable point, and away from the solvable point we have calculated the signal from the channel with $E_{\mu} = E_{\nu} = E_{\lambda}$, which is the channel expected to give rise to signatures of deconfinement. We have also verified that the contribution from the intermediate channel with $E_{\nu} = E_{\mu} + E_{\lambda}$ does not cancel the signal of interest at the solvable point, where this extra (more complicated) channel can be analytically treated. Thus, linear response on X-cube can tell that a local field creates one dimensional excitations (lineons), through DOS. 2DCS can then verify that the lineons are deconfined. 
 
\subsubsection{$(\alpha, \beta, \gamma) = (z,z,z)$} \label{sssec:planon}
 
\begin{figure*}[t]
\centering
\includegraphics[width=\linewidth]{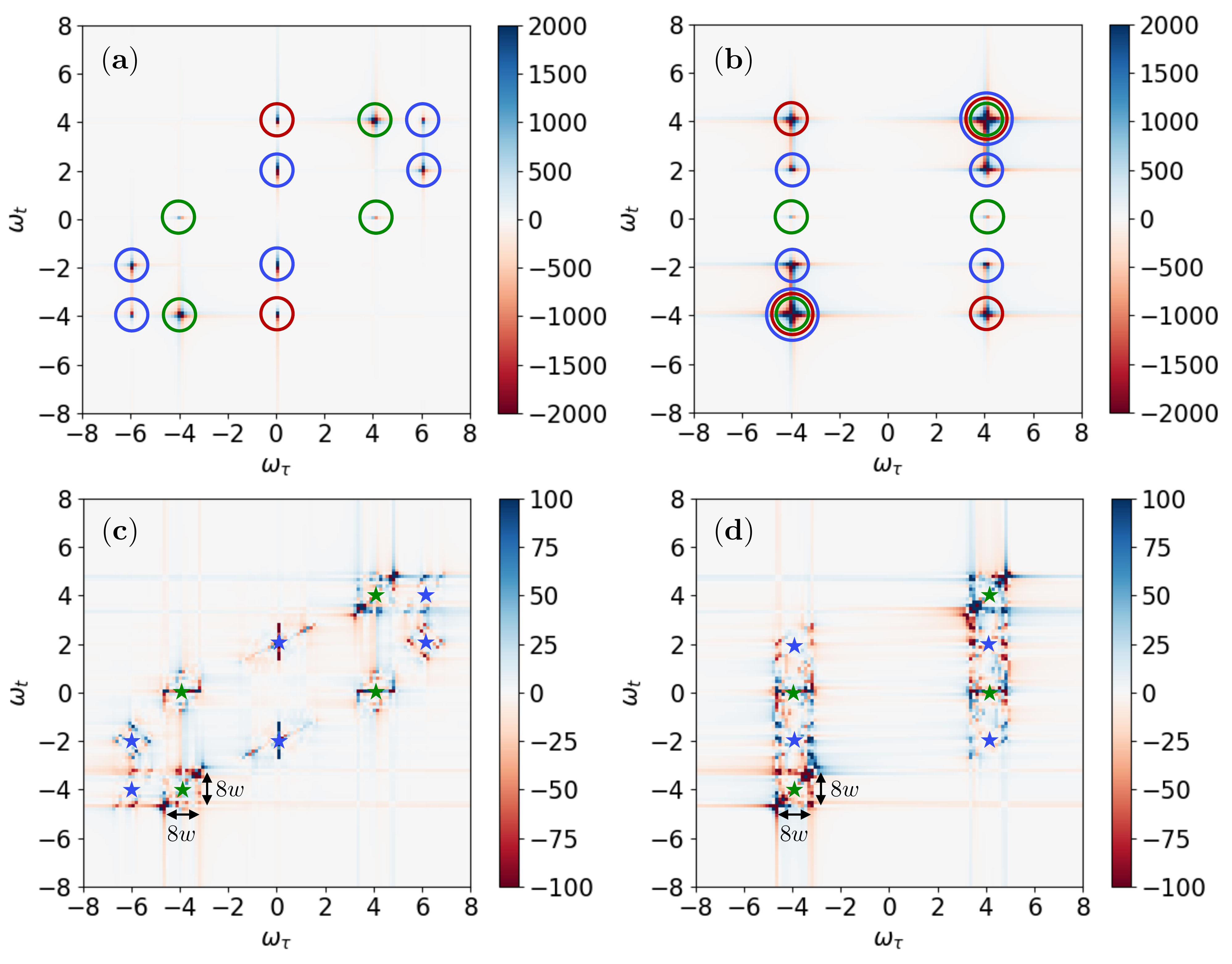}
\caption{Imaginary part of 2D Fourier transform $\mathcal{F}_{t,\tau}$ of the third-order susceptibilities $\chi_{zzzz}$ for X-cube model ($K=1$) (a, b) without perturbations and (c, d) with generic perturbations.
(a) $\mathrm{Im}\,\mathcal{F}_{t,\tau}[\chi_{zzzz}(t,t+\tau,t+\tau)]$ and (b) $\mathrm{Im}\,\mathcal{F}_{t,\tau}[\chi_{zzzz}(t,t,t+\tau)]$
of X-cube model are qualitatively similar to those of Ising model except the signals marked with green circles.
The green circled signals are coming from free motion of a pair of planons having restricted mobility within two-dimensional planes ($E_\mu=E_\nu=E_\lambda$ process in Sec.~\ref{sssec:planon}).
Again, the signals at $(\omega_\tau, \omega_t) = (\pm 4, 0)$ suggest deconfined excitations of X-cube model.
Red circled signals are due to quantum processes involving four planons ($E_\nu = E_\mu + E_\lambda$ process in Sec.~\ref{sssec:planon}), and blue circled signals are originating from the process with two planons and two immobile fractons ($E_\nu = E_\mu + 2K = E_\lambda + 2K$ process in Sec.~\ref{sssec:planon}).
(c) $\mathrm{Im}\,\mathcal{F}_{t,\tau}[\chi_{zzzz}(t,t+\tau,t+\tau)]$ and (d) $\mathrm{Im}\,\mathcal{F}_{t,\tau}[\chi_{zzzz}(t,t,t+\tau)]$ of the perturbed X-cube model.
The nonlinear signals from the dispersive two-planon ($E_\mu=E_\nu=E_\lambda$) and two-planon-two-fracton ($E_\nu = E_\mu + 2K = E_\lambda + 2K$) processes are shown.
Recall that the blue circled signals of Ising model (FIG.~\ref{fig: Ising2dcs}) are coming from a \textit{single} zero momentum excitation of the two-magnon bound state.
Hence, they do not show significant spreading in the presence of perturbations.
However, the blue circled signals of X-cube model are originating from nonlinear dynamics of two finite momentum carrying planons scattering with two immobile fractons.
Thus, we can see clear spread of the signals by the full bandwidth of the single planon dispersion $4w = 0.8$. Blue and green stars mark locations of the signals without the perturbations.}
\label{fig:planon}
\end{figure*}

  Now $\mu$ and $\lambda$ both contain a pair of planeons with zero net momentum. Since the planeons are mobile there is an associated continuum of energies. Meanwhile options for $\nu$ include zero planeons ($E_{\nu}=0$), one pair of planeons ($E_{\mu} = E_{\nu} = E_{\lambda}$), two pairs of planeons ($E_{\nu} = E_{\mu} + E_{\lambda}$). Thus far the analysis parallels the toric code case, with planeons taking the place of visons. In particular, there will be a diagonal streak in the two dimensional Fourier transform, with no offset, as well as a streak along the $\omega_{\tau}$ axis with no offset, as a signature of deconfined fractionalized excitations which are their own antiparticles. However, there is one additional intermediate option, whereby $\nu$ contains one pair of planeons {\bf and also} two fractons (which each have zero momentum), such that $E_{\mu} = E_{\lambda}$ but $E_{\nu} = E_{\mu}+2K$. Lets explore the consequences of this additional channel, bearing in mind that $E_{\mu}$ can take a continuum of values (outside the two planeon gap), and that we are only interested in features that are sharp in at least one direction. 
  
  The first and sixth terms in Eq.\ref{chi31} will give rise to diagonal streaks with  offset $2K$ in the $\omega_t$ direction, outside the two planeon gap, while the second and fifth terms will give rise to streaks along $\omega_{\tau}$ with  offset $2K$ in the $\omega_t$ direction outside the two planeon gap. Meanwhile the first and sixth terms in Eq.\ref{chi32} will again produce diagonal streaks with no offset, while the middle four terms in Eq.\ref{chi32} generate streaks in the $\omega_{\tau}$ direction with offset $2K$ in the $\omega_t$ direction outside the planeon gap. Thus, the main `new' consequence is offset stripes. Now, offset stripes also appeared in the TFIM, but there the offset was equal to the linear response gap. Here the offset is equal to {\it half} the linear response gap. Altogether this is quite informative. The lack of any sharp features in linear response (besides the gap) tells us that local fields do not create individual local sharp quasiparticles. The appearance of vertical streaks along axis in the two dimensional Fourier transform tell us that local fields {\it do} create deconfined pairs of quasiparticles (planeons). The additional appearance of {\it sharp} streaks at an offset a fraction of the linear response gap tells us that the planeons can further fractionalize, but also that the objects generated by planeon fractionalization are non-dispersive (fractons). Altogether 2DCS provides a crisp diagnostic for planeons, fractons, and lineons in the X-cube phase. 
  
  In the Appendix, we have provided explicit calculations for the X-cube model in the z polarization. Calculations are exact at the solvable point, whereas away from the solvable point they include the channels with $E_{\nu} = E_{\mu} = E_{\lambda}$ and $E_{\nu} = E_{\mu} + 2K$, which produce the key signals of interest. Figure \ref{fig:planon} plots the results. 



 \subsection{Haah code}
We will consider here only the $xxx$ experimental geometry - the $zzz$ geometry has analogous behavior. 
Now both $|\mu\rangle$ and $|\lambda\rangle$ are produced by acting on the ground state with a singe $X$ operator, and hence contain a `tetrahedron' of four fractons, with energy $E_{\mu} = E_{\nu} = 4$. This cannot be subdivided into two (or more) mobile excitations, so given the constraint that $|\mu\rangle$ and $|\lambda\rangle$ have zero momentum, it follows that they also have a sharply defined energy, rather than a continuum. What about $|\nu\rangle$? This can contain either: zero fractons ($E_{\nu}=0)$), six fractons ($E_{\nu} = 6 + \delta'$), or eight fractons ($E_{\nu} = 8 + 2E_k)$. In the last we have taken account of the fact that composites of four fractons are locally creatable and hence mobile objects, which can carry momentum. 

Let us compare to the Ising model in three dimensions (i.e. the same spatial dimensionality as the Haah code). In the Ising model, we have $E_{\mu}=E_{\nu} = 6$ (equal to the linear response gap) and $E_{\nu} = 0$ or $E_{\nu} = 12+2E_k$ (2 spin flips with equal and opposite momentum), or $E_{\nu} = 10+\delta'$ (bound state of two adjacent spin flips). While there is a quantitative difference to the Haah code, there is not a qualitative difference - each time we act with an $X$ operator we create a single mobile excitation (a single spin flip in the Ising model, a tetrad of excited cubes in the Haah code), and two adjacent mobile excitations can form a bound state. While the quantitative differences are suggestive that what is going on in the Haah code may be different, there is not therefore an unambiguous diagnostic in 2DCS. 

It should however be possible to diagnose the Haah code in higher order non-linear response. A key property of the Haah code is that acting with an appropriate fractal membrane of $X$ operators only produces four cube excitations at the `corners' \cite{fracton2}. Thus, while acting with a single $X$ operator produces a state with energy equal to the linear response gap, acting with four $X$ operators in a tetrahedral arrangement also produces a state with the same energy gap. In contrast, in the Ising model there is no way to produce a state with energy equal to the linear response gap by acting with four operators. If we go to sufficiently high order to access this intermediate state, then we should be able to {\it qualitatively} distinguish the Haah code from the trivial case of the Ising model. However, this will require consideration of at least the seventh order non-linear response, which is beyond the scope of the present paper. Experimentally also, measurement of a seventh order non-linear response may be challenging - although it could certainly be accomplished in principle, e.g. with a seven pulse experiment. Unambiguous diagnosis of the Haah code is therefore a more challenging task than identification of its conventional spin liquid or type-I fracton counterparts. 
 \section{Conclusions}
 \label{sec: conclusions}
Thus, we have identified purely spectroscopic fingerprints of both conventional gapped spin liquids and gapped fractonic phases using the technique of 2DCS spectroscopy. The easiest to diagnose are type I fracton phases with lineon (one dimensional) excitations - these can be diagnosed even in linear response through the existence of one dimensional Van Hove singularities in the absorbtion spectrum. If we turn to non-linear response and the 2DCS spectrum, then type I fracton phases have further fingerprints of deconfinement, of the existence of lineon excitations, and also of the existence of totally immobile fracton excitations. Conventional gapped spin liquids are intermediate in subtlety to detect. In linear response there are no sharp features besides the bulk gap. However in non-linear response and the 2DCS spectrum, there appear sharp signatures of the existence of deconfined quasiparticles. Finally, type II fracton phases (such as the Haah code) are most subtle to diagnose. Linear response has no clear signatures, and even 2DCS gives only quantitative (but not qualitative) distinctions from trivial possibilities.  An unambiguous diagnosis of the Haah code likely requires a consideration of high order non-linear response (seventh order susceptibility should suffice), which may be challenging to access experimentally.

A number of future directions present themselves for consideration. For instance, thus far we have considered only the simplest version of the experiment, where each of the pulses (and the observed signal) has the same polarization. Crossed polarizations would be interesting to consider in future work, and could yield additional information beyond the `single polarization' experiments discussed herein. Additionally, the calculations presented herein have been for a system prepared in the ground state. Extension to low temperature Gibbs states would also be a natural problem for future work.   Perhaps most significantly, we have thus far ignored dissipative line broadening. However, dissipation will inevitably be present, whether extrinsic (due to coupling to e.g. phonons), or intrinsic (due to decay to the multi-excitation continuum). This will produce line broadening, and a quantitative understanding thereof could be a fruitful endeavour. Indeed, it seems plausible \cite{premhaahnandkishore} that a careful analysis of the temperature dependence of line broadening might itself provide a clean diagnostic for otherwise difficult to detect phases like the Haah code. The ability of 2DCS to distinguish between energy relaxation ($T_1$ time) and dephasing ($T_2$ time) may also be useful in this regard. 

We have thus far also assumed that we are dealing with {\it clean} systems, whereas realistic experimental systems will inevitably be disordered. Disorder is not expected to be important for the signals discussed herein, but would certainly be important for any analysis of dissipative lineshapes, where the ability of 2DCS to distinguish `intrinsic' line broadening from inhomogenous broadening could be particularly useful. Incorporating disorder into the analysis would thus also be a fruitful project for the future. 

Furthermore, thus far we have identified sharp signatures of deconfinement, lineons, fractons etc, but while these serve as crisp diagnostics for a fractionalized phase (or a fracton phase), they do not establish {\it which} fractionalized (or fractonic) phase we are dealing with. For example, the diagnostics we identified would not be able to distinguish between the $Z_2$ spin liquid represented by the toric code and that represented by the doubled semion model \cite{doubledsemion}, nor between the X-cube phase and the semionic X-cube phase \cite{semionxcube}. Identifying diagnostics able to unambiguously identify which phase within each class we were dealing with would also be a fruitful topic for future work. Similarly, diagnostics capable of identifying symmetry enriched phases, akin to \cite{EssinHermele} would also be worth identifying. 

Finally, thus far we have considered only the third order non-linear response, probed in a two-pulse experiment with 2DCS. There are other possibilities. For instance, one could consider the third order response probed through a three pulse experiment, and analyzed via the {\it three} dimensional Fourier transform, or we could consider fifth (or higher) order nonlinear susceptibilities. Indeed we have argued that the seventh order non-linear response should be particularly interesting, as it likely offers a clean diagnostic for the Haah code. All these would no doubt yield additional information, at the cost of complicating the necessary experiment. Regardless, it appears clear that the multidimensional spectroscopy technique places a powerful new tool at our disposal, which can be used to identify exotic phases that would be difficult to diagnose via conventional techniques.  
 
 {\bf Acknowledgements} We acknowledge useful conversations with N.P. Armitage. RN is supported by the Department of Energy under award number DE-SC0021346. WC and YBK are supported by the NSERC of Canada and the Center for Quantum Materials at the University of Toronto. YBK is also supported by the Killam Research Fellowship from the Canada Council for the Arts. Numerical
 computations were performed on the Cedar and Niagara clusters, which are hosted by WestGrid and SciNet in partnership with ComputeCanada.
\bibliography{library}

\clearpage
\onecolumngrid
\setcounter{equation}{0}
\setcounter{figure}{0}
\setcounter{table}{0}
\setcounter{page}{1}
\setcounter{section}{0}
\setcounter{subsection}{0}

\pagebreak
\widetext
\begin{center}
\textbf{\large Supplemental Materials for ``Spectroscopic fingerprints of gapped quantum spin liquids, both conventional and fractonic"}
\end{center}
\setcounter{equation}{0}
\setcounter{figure}{0}
\setcounter{table}{0}
\setcounter{page}{1}
\makeatletter
\renewcommand{\theequation}{S\arabic{equation}}
\renewcommand{\thefigure}{S\arabic{figure}}

\section{Nonlinear susceptibilities of the stabilizer codes}

In this appendix, we provide detailed calculations for the nonlinear susceptibilities of toric code and X-cube model.
Since we can find the exact energy eigenvalues and eigenstates of the models, the dynamical four-point correlation functions [Eq.~(\ref{eq:R})] can be exactly calculated.
In the presence of small perturbations, the model is no longer exactly solvable.
However, we can make further analysis by introducing appropriate dispersion to the elementary excitations in the plane wave basis.

We first note that the unperturbed stabilizer codes have nonvanishing nonlinear susceptibilities $\chi_{\gamma \alpha \alpha\beta}(t, t+\tau, t+\tau)$ and $\chi_{\gamma \alpha\beta\beta}(t,t,t+\tau)$ only if $\alpha=\beta=\gamma$.
Since different Pauli operators create/annihilate different types of excitations, nonvanishing correlation functions must have even number of each type of Pauli operators on a cubic lattice.
For example, $\langle ZZZZ \rangle$ and $\langle XXZZ\rangle$ can be (not necessarily but possibly) finite, but correlation functions like $\langle XZZZ \rangle$ must be zero because excitations created by $X$ cannot be annihilated by $Z$.
Therefore $\chi_{\gamma \alpha \alpha\beta}(t, t+\tau, t+\tau) = 0$ when $\gamma \neq \beta$ and $\chi_{\gamma \alpha \beta\beta} (t,t,t+\tau) = 0$ when $\gamma \neq \alpha$.

From the explicit calculations, one can further confirm that $\chi_{\beta \alpha \alpha\beta}(t, t+\tau, t+\tau) = \chi_{\alpha \alpha \beta\beta} (t,t,t+\tau) = 0$ if $\alpha \neq \beta$.
Although the matrix elements $S$ are finite, the quantum phases destructively interfere so that all terms in Eqs.~(\ref{chi31}) and (\ref{chi32}) add up to zero.
This cancellation has to happen because the nonlinear susceptibility should be independent of the system size [Eq.~(\ref{eq:chidef})] while the matrix elements $S \propto L^{2d}$ if $\alpha \neq \beta$.
Since the dynamics of different types of excitations are completely decoupled, creation and annihilation of each type of excitation gives factor of $N$ to the matrix element $S$.
For example,
\begin{align}
S_{zzxx}(\mu,\nu,\lambda)
&= \langle 0 |M_z |\mu \rangle \langle \mu | M_z  | \nu \rangle
\langle \nu | M_x | \lambda \rangle \langle \lambda | M_x | 0 \rangle
= \sum_{j,k=1}^N  \langle 0 |Z_j |\mu \rangle \langle \mu | Z_j  | \nu \rangle
\langle \nu | X_k | \lambda \rangle \langle \lambda | X_k | 0 \rangle \\
&=\left( \sum_{j=1}^N \langle 0 |Z_j |\mu \rangle \langle \mu | Z_j  | \nu \rangle \right) \left( \sum_{k=1}^N  \langle \nu | X_k | \lambda \rangle \langle \lambda | X_k | 0 \rangle \right)
\propto N^2 \propto L^{2d},
\end{align}
where $N$ is the total number of spins on a lattice.
As $\chi \propto \frac{1}{L^d} S \propto L^d \neq \mathcal{O}(1)$ is inconsistent with the definition of the nonlinear susceptibility, we can expect that the susceptibility must be vanishing when $\alpha \neq \beta$.

With the nonvanishing condition $\alpha = \beta =\gamma$, Eqs.~(\ref{chi31}) and (\ref{chi32}) can be simplified as
\begin{align}
\chi_{\alpha\alpha\alpha\alpha}(t, t+\tau, t+\tau) \equiv \chi^{(a)}_\alpha(t,\tau) &= \frac{2}{N} \sum_{\mu\nu\lambda} \mathrm{Im} \left[\mathcal{R}^{(a)}(\mu,\nu,\lambda;t,\tau)
\langle 0 |M_\alpha |\mu \rangle \langle \mu | M_\alpha  | \nu \rangle
\langle \nu | M_\alpha | \lambda \rangle \langle \lambda | M_\alpha | 0 \rangle \right], \label{eq:chi3a} \\
\chi_{\alpha\alpha\alpha\alpha}(t, t+\tau, t+\tau) \equiv \chi^{(b)}_\alpha(t,\tau) &= \frac{2}{N} \sum_{\mu\nu\lambda}
\mathrm{Im} \left[ \mathcal{R}^{(b)}(\mu,\nu,\lambda;t,\tau)
\langle 0 |M_\alpha |\mu \rangle \langle \mu | M_\alpha  | \nu \rangle
\langle \nu | M_\alpha | \lambda \rangle \langle \lambda | M_\alpha | 0 \rangle \right], \label{eq:chi3b}
\end{align}
where
\begin{align}
\mathcal{R}^{(a)}(\mu,\nu,\lambda;t,\tau) = 2 e^{i(E_\mu-E_\lambda)\tau}e^{i(E_\nu - E_\lambda)t}
+ e^{i(E_\nu - E_0)\tau}e^{i(E_\nu-E_\lambda)t}+e^{i(E_0-E_\nu)\tau}e^{i(E_0-E_\mu)t}, \label{eq:Ra}\\
\mathcal{R}^{(b)}(\mu,\nu,\lambda;t,\tau) = e^{i(E_0 - E_\lambda)\tau}e^{i(E_\nu - E_\lambda)t}
+ 2e^{i(E_\mu-E_0)\tau}e^{i(E_\nu-E_\lambda)t} + e^{i(E_0 - E_\lambda)\tau}e^{i(E_0-E_\mu)t},\label{eq:Rb}
\end{align}
and $E_{\mu}, E_{\nu}, E_{\lambda}$ and $|\mu\rangle, |\nu\rangle, |\lambda\rangle$ are the energy eigenvalues and eigenstates of the stabilizer codes.

\subsection{2D Ising ferromagnet}

We use 2D ferromagntic Ising model as a reference model for conventional magnetic systems.
The model has ferromagnetic ground state and each spin flip ($X$ operator) results in a gapped magnon excitation.
On a square lattice, a single spin flip has energy cost $\varepsilon = 4$, and non-neighbouring two spin flips cost $\varepsilon = 8$. When two neighbouring spins are flipped, then the energy cost is $\varepsilon =6$.

By plugging in the energy cost into Eqs.~(\ref{eq:chi3a}) and (\ref{eq:chi3b}), we can calculate the third order susceptibilities $\chi_x^{(a/b)}(t,\tau)$ for Ising model on a square lattice:
\begin{align}
\chi_x^{(a/b)}(t,\tau) &=
\frac{2}{L^2}\,\Theta(t)\Theta(\tau) \sum_{l,l'}
\sum_{\mu\lambda}
\mathrm{Im}\left[ \mathcal{R}^{(a/b)}(\mu,0,\lambda;t,\tau) 
\langle 0 | X_l | \mu \rangle \langle \mu | X_l | 0 \rangle
\langle 0 | X_{l'} | \lambda \rangle \langle \lambda | X_{l'} | 0 \rangle \right] \\
&+\frac{2}{L^2} \Theta{(t)}\Theta{(\tau)} \sum_{l} \sum_{\eta = \pm \hat{x},\pm \hat{y}} \sum_{\mu\nu\lambda}
\mathrm{Im}\left[ \mathcal{R}^{(a/b)}(\mu,\nu,\lambda;t,\tau) 
\langle 0 | X_l | \mu \rangle \langle \mu | X_{l+\eta} | \nu \rangle
\langle \nu | X_{l+\eta} | \lambda \rangle \langle \lambda | X_{l} | 0 \rangle \right] \\
&+\frac{2}{L^2} \Theta{(t)}\Theta{(\tau)} \sum_{l} \sum_{\eta = \pm \hat{x},\pm \hat{y}} \sum_{\mu\nu\lambda}
\mathrm{Im}\left[ \mathcal{R}^{(a/b)}(\mu,\nu,\lambda;t,\tau) 
\langle 0 | X_l | \mu \rangle \langle \mu | X_{l+\eta} | \nu \rangle
\langle \nu | X_l | \lambda \rangle \langle \lambda | X_{l+\eta} | 0 \rangle \right]  \\
&+ \frac{2}{L^2}  \Theta{(t)}\Theta{(\tau)} \sum_{l} \sum_{l' \neq l, l\pm \hat{x}, l\pm \hat{y}} \sum_{\mu\nu\lambda}
\mathrm{Im}\left[ \mathcal{R}^{(a/b)}(\mu,\nu,\lambda;t,\tau) 
\langle 0 | X_l | \mu \rangle \langle \mu | X_{l'} | \nu \rangle
\langle \nu | X_{l'} | \lambda \rangle \langle \lambda | X_l | 0 \rangle \right] \\
&+ \frac{2}{L^2}  \Theta{(t)}\Theta{(\tau)} \sum_{l} \sum_{l' \neq l, l\pm \hat{x}, l\pm \hat{y}} \sum_{\mu\nu\lambda}
\mathrm{Im}\left[ \mathcal{R}^{(a/b)}(\mu,\nu,\lambda;t,\tau) 
\langle 0 | X_l | \mu \rangle \langle \mu | X_{l'} | \nu \rangle
\langle \nu | X_{l} | \lambda \rangle \langle \lambda | X_{l'} | 0 \rangle \right] \\
&=2L^2\,\Theta(t)\Theta(\tau) \mathrm{Im}\, \mathcal{R}^{(a/b)}(\varepsilon_\mu=4,\varepsilon_\nu=0,\varepsilon_\lambda=4;t,\tau) \\
&+ 16 \,\Theta(t)\Theta(\tau)  \mathrm{Im}\,\mathcal{R}^{(a/b)}(\varepsilon_\mu=4,\varepsilon_\nu = 6,\varepsilon_\lambda=4;t,\tau) \\
&+ 4(L^2-5) \,\Theta(t)\Theta(\tau) \mathrm{Im}\, \mathcal{R}^{(a/b)}(\varepsilon_\mu=4,\varepsilon_\nu = 8,\varepsilon_\lambda=4;t,\tau) \\
&=2L^2 \,\Theta(t)\Theta(\tau) \mathrm{Im}\left[ \mathcal{R}^{(a/b)}(\varepsilon_\mu=4,\varepsilon_\nu=0,\varepsilon_\lambda=4;t,\tau) 
+ 2 \mathcal{R}^{(a/b)}(\varepsilon_\mu=4,\varepsilon_\nu = 8,\varepsilon_\lambda=4;t,\tau)  \right] \\
&+4\,\Theta(t)\Theta(\tau) \mathrm{Im}\left[4 \mathcal{R}^{(a/b)}(\varepsilon_\mu=4,\varepsilon_\nu = 6,\varepsilon_\lambda=4;t,\tau) - 5\mathcal{R}^{(a/b)}(\varepsilon_\mu=4,\varepsilon_\nu = 8,\varepsilon_\lambda=4;t,\tau) \right]. \label{eq:Ising_realspace_general}
\end{align}
The terms proportional to $L^2$ are precisely cancelled, which is consistent with the nonextensive definition of the nonlinear susceptibilities.
Then
\begin{align}
\chi_x^{(a)}(t,\tau)&= 8\,\Theta(t)\Theta(\tau) \left[ 4\sin 2t - 5 \sin 4t - 2\sin (4t+6\tau) +2\sin(2t+6\tau) \right],\\
\chi_x^{(b)}(t,\tau)&= 4\,\Theta(t)\Theta(\tau) \left[ 4\sin(2t-4\tau) +8\sin(2t+4\tau) -5\sin 4(t-\tau) - 9\sin 4(t+\tau) \right].
\end{align}

Under generic perturbations, a single spin-flip excitation (magnon) becomes dispersive.
Since the contribution from perturbative process involving $|\nu\rangle = |0\rangle$ is eventually cancelled, we only need to consider nonlinear dynamics of two neighboring spin flip composite ($\varepsilon_\nu = 6 + \delta$) and dispersive motion of two separated magnons ($\varepsilon_\nu = 8 + \varepsilon_{q_1} + \varepsilon_{q_2}$), where $\delta$ and $\varepsilon_{q_{1,2}}$ are perturbative corrections to the static energy cost.
%

Let's first consider the process with two separated magnons.
With the resolution of identity in the plane wave basis $|p\rangle = a_p^\dagger |0\rangle = \frac{1}{L} \sum_\xi e^{i p \cdot \xi} a_\xi^\dagger|0\rangle$,
\begin{multline}
\chi_x^{(a/b)}(t,\tau) = \frac{2}{L^2}\,\Theta(t)\Theta(\tau)\,\mathrm{Im}\left[
\frac{1}{2} \sum_{p}\sum_{q_1,q_2}\sum_r \mathcal{R}^{(a/b)}(p,q_1,q_2,r;t,\tau)
\sum_{jklm}
\langle 0 | X_j | p \rangle \langle p | X_k |q_1, q_2 \rangle \langle q_1, q_2 | X_l | r\rangle
\langle r| X_m |0\rangle \right].
\end{multline}

The matrix element is calculated using the standard Holstein-Primakoff transformation,
\begin{align}
X_j &= \sqrt{1-a_j^\dagger a_j} a_j + a_j^\dagger \sqrt{1-a_j^\dagger a_j}.
\end{align}
Then,
\begin{align}
\sum_j \langle 0 | X_j |p\rangle
&= \frac{1}{L} \sum_j \sum_\xi e^{ip\cdot \xi} \langle 0| \left( \sqrt{1-a_j^\dagger a_j} a_j + a_j^\dagger \sqrt{1-a_j^\dagger a_j} \right)
a_\xi^\dagger |0\rangle
= \frac{1}{L}\sum_j \sum_\xi e^{ip\cdot \xi} \langle 0|  \sqrt{1-a_j^\dagger a_j} a_j a_\xi^\dagger |0\rangle \\
&=\frac{1}{L}\sum_{j,\xi} e^{ip\cdot \xi} \delta_{j\xi} = \sum_{j} e^{ip\cdot R_j} = L \delta_{p,0}. \label{eq:p0}
\end{align}
For the matrix element with the two-magnon state, the restricted summation $\textstyle \sideset{}{'}\sum$ carefully exclude the case where two spin-flips are right next to each other:
\begin{align}
\sum_k \langle p | X_k | q_1, q_2\rangle &=
\frac{1}{L^3} \sum_k \sum_{\xi} \sideset{}{'}\sum_{\zeta_1, \zeta_2} e^{-i p \cdot \xi +iq_1 \cdot \zeta_1 +iq_2 \cdot \zeta_2}
\langle 0 | a_\xi  \left( \sqrt{1-a_k^\dagger a_k} a_k + a_k^\dagger \sqrt{1-a_k^\dagger a_k} \right)
a_{\zeta_1}^\dagger a_{\zeta_2}^\dagger |0\rangle \\
&= \frac{1}{L^3} \sum_k \sum_{\xi} \sideset{}{'}\sum_{\zeta_1, \zeta_2}  e^{-i p \cdot \xi +iq_1 \cdot \zeta_1 +iq_2 \cdot \zeta_2}
\langle 0 | a_\xi \sqrt{1-a_k^\dagger a_k} a_k a_{\zeta_1}^\dagger a_{\zeta_2}^\dagger |0\rangle \\
&= \frac{1}{L^3} \sum_k \sum_{\xi \neq k}\sideset{}{'} \sum_{\zeta_1, \zeta_2}
 e^{-i p \cdot \xi +iq_1 \cdot \zeta_1 +iq_2 \cdot \zeta_2}
\langle 0 | a_\xi a_k a_{\zeta_1}^\dagger a_{\zeta_2}^\dagger |0\rangle  \\
&=\frac{1}{L^3} \sum_k \sum_{\xi \neq k}\sideset{}{'} \sum_{\zeta_1, \zeta_2}
 e^{-i p \cdot \xi +iq_1 \cdot \zeta_1 +iq_2 \cdot \zeta_2}
\left(\delta_{\zeta_1,\xi}\delta_{\zeta_2, k}+\delta_{\zeta_1,k}\delta_{\zeta_2, \xi} \right)  \\
&= \frac{1}{L^3} \sum_k \sum_{\xi \neq k, k \pm\hat{x}, k \pm \hat{y}} e^{i(q_1 - p)\cdot \xi + i q_2 \cdot R_k} + e^{i(q_2 - p)\cdot \xi + i q_1 \cdot R_k} \\
&= \frac{1}{L^3} \sum_k \sum_{\xi} e^{i(q_1 - p)\cdot \xi + i q_2 \cdot R_k} + e^{i(q_2 - p)\cdot \xi + i q_1 \cdot R_k} \nonumber \\
&- \frac{1}{L^3} \sum_k e^{i(q_1 + q_2 - p) \cdot R_k} \left[2 + \sum_{\eta = \pm\hat{x},\pm\hat{y}}e^{i(q_1 - p)\cdot \eta} 
+ e^{i(q_2 - p)\cdot \eta }\right] \\
&=L \left( \delta_{q_1,p}\delta_{q_2,0} +  \delta_{q_1,0}\delta_{q_2,p} \right)
-\frac{2}{L} \delta_{q_1 + q_2, p} \left(1 + \cos q_{1x} + \cos q_{1y} + \cos q_{2x} + \cos q_{2y} \right)
\end{align}
Due to the Kronecker delta in Eq.~(\ref{eq:p0}), the only relevant matrix elements are
\begin{align}
\sum_k \langle 0 | X_k | q, -q\rangle = 2L \delta_{q,0} -\frac{2}{L} \left(1 + 2\cos q_{x} + 2\cos q_{y} \right).
\end{align}

When two spin flips are neighbouring, the energy cost is smaller than the cost for two separated magnons. So we treat this case separately.
Since two magnons are moving altogether, both magnons have the same momentum.
So the relevant matrix element is
\begin{align}
\sum_k \langle p | X_k | q,q \rangle &= \frac{1}{L^2}
\sum_k \sum_{\xi,\zeta} \sum_{\eta=\hat{x},\hat{y}} e^{-ip\cdot \xi + i q\cdot(2\zeta+\eta)}
\langle 0 | a_\xi \left( \sqrt{1-a_k^\dagger a_k} a_k + a_k^\dagger \sqrt{1-a_k^\dagger a_k} \right)
a_{\zeta}^\dagger a_{\zeta + \eta}^\dagger |0\rangle \\
&=\frac{1}{L^2}\sum_k \sum_{\xi\neq k}\sum_\zeta \sum_{\eta=\hat{x},\hat{y}} e^{-ip\cdot \xi + i q\cdot(2\zeta+\eta)}
\langle 0 | a_\xi a_k a_{\zeta}^\dagger a_{\zeta + \eta}^\dagger |0\rangle \\
&=\frac{1}{L^2}\sum_k \sum_{\xi\neq k}\sum_\zeta \sum_{\eta=\hat{x},\hat{y}} e^{-ip\cdot \xi + i q\cdot(2\zeta+\eta)}
(\delta_{\zeta,\xi}\delta_{\zeta+\eta,k} + \delta_{\zeta,k}\delta_{\zeta+\eta,\xi}) \\
&=\frac{1}{L^2} \sum_k  e^{i(2q-p)\cdot R_k}\sum_{\eta=\hat{x},\hat{y}} \left[ e^{-i(q-p)\cdot \eta} +e^{i(q-p)\cdot \eta} \right] =2 \delta_{2q,p} (\cos q_x + \cos q_y).
\end{align}
Due to momentum conservation constraint imposed by the Kronecker deltas, we only need to consider the case $p = q = 0$.

If we add up those two contributions,
\begin{align}
\chi_x^{(a/b)}(t,\tau) &=4\,\Theta(t)\Theta(\tau)\,\mathrm{Im}\left[
\mathcal{R}^{(a/b)}(0;t,\tau) (L^2 - 10 )
+ \frac{1}{L^2}\sum_q \mathcal{R}^{(a/b)}(q;t,\tau)(1+2\cos q_x + 2\cos q_y)^2
\right] \\
&+16\,\Theta(t)\Theta(\tau) \, \mathrm{Im}\, \widetilde{\mathcal{R}}^{(a/b)}(0;t,\tau) \\
&= 4 L^2 \,\Theta(t)\Theta(\tau) \,\mathrm{Im}\,\mathcal{R}^{(a/b)}(0;t,\tau)\label{eq:Ising_8_extensive} \\
&+16\,\Theta(t)\Theta(\tau) \,\mathrm{Im}\,\widetilde{\mathcal{R}}^{(a/b)}(0;t,\tau) \label{eq:Ising_6} \\
&+4 \,\Theta(t)\Theta(\tau) \left[\frac{1}{L^2}\sum_q  (1+2\cos q_x + 2\cos q_y)^2\,
\mathrm{Im}\,\mathcal{R}^{(a/b)}(q;t,\tau) - 10\,\mathrm{Im}\,\mathcal{R}^{(a/b)}(0;t,\tau)\right], \label{eq:Ising_8_dispersive}
\end{align}
where $\mathcal{R}^{(a/b}(q;t,\tau) \equiv \mathcal{R}^{(a/b)}(p=0,q, -q,r=0;t,\tau)$ and  $\widetilde{\mathcal{R}}^{(a/b)}(0;t,\tau)$are
\begin{align}
\mathcal{R}^{(a)}(q;t,\tau) &=e^{- i \varepsilon_0 t} \left[ 2 e^{i \varepsilon_q t }
+ e^{i \varepsilon_q (t+\tau)}+e^{-i\varepsilon_q \tau} \right], \\
\mathcal{R}^{(b)}(q;t,\tau) &= e^{- i\varepsilon_0 (t+\tau)}e^{i \varepsilon_q t}
+ 2e^{-i\varepsilon_0 (t-\tau)}e^{i\varepsilon_q t} + e^{-i\varepsilon_0(t+ \tau)}, \\
\widetilde{\mathcal{R}}^{(a)}(0;t,\tau) &=e^{- i \varepsilon_0 t} \left[ 2 e^{i \widetilde{\varepsilon}_0 t }
+ e^{i \widetilde{\varepsilon}_0 (t+\tau)}+e^{-i\widetilde{\varepsilon}_0 \tau} \right], \\
\widetilde{\mathcal{R}}^{(b)}(0;t,\tau) &= e^{- i\varepsilon_0 (t+\tau)}e^{i \widetilde{\varepsilon}_0 t}
+ 2e^{-i\varepsilon_0 (t-\tau)}e^{i\widetilde{\varepsilon}_0 t} + e^{-i\varepsilon_0(t+ \tau)},
\end{align}
with $\varepsilon_0 = 4 + w(\cos 0 + \cos 0) = 4 + 2w$, $\varepsilon_q = 8 + 2w(\cos q_x + \cos q_y)$, and $\widetilde{\varepsilon}_0 = 6 + 4w$.

The extensive term, Eq.~(\ref{eq:Ising_8_extensive}), is responsible for the cancellation of the other extensive term from the process involving $|\nu\rangle = |0\rangle$.
Hence, we only take the system size independent terms.
Note that the expression [Eq.~(\ref{eq:Ising_6}) + (\ref{eq:Ising_8_dispersive})] reproduces the exact susceptibility [Eq.~(\ref{eq:Ising_realspace_general})] without perturbations (i.e., $w = 0$).

\subsection{2D Toric code}

Magnetic fields along $x$ and $z$-axis excite $e$ (vertex excitation, \textit{i.e.}, bosonic charge) and $m$ (plaquette excitation, \textit{i.e.}, $\mathbb{Z}_2$ flux) particles of the two-dimensional toric code, respectively. $y$-polarized fields excite both $e$ and $m$ particles.
On a square lattice, dynamics of the fluxes ($\chi_{xxxx}$) and the bosonic charges ($\chi_{zzzz}$) are the same.
So let's focus on the nonlinear response of the bosonic charges under two $z$-polarized incident pulses:
\begin{align}
\chi_z^{(a/b)}(t,\tau) &= \frac{1}{L^2} \Theta{(t)}\Theta{(\tau)} \sum_{l,l'} \sum_{\mu\lambda}
\mathrm{Im}\left[ \mathcal{R}^{(a/b)}(\mu,0,\lambda;t,\tau) 
\langle 0 | Z_l | \mu \rangle \langle \mu | Z_l | 0 \rangle
\langle 0 | Z_{l'} | \lambda \rangle \langle \lambda | Z_{l'} | 0 \rangle \right] \label{eq:tc_nu0} \\
&+ \frac{1}{L^2}  \Theta{(t)}\Theta{(\tau)} \sum_{l \neq l'} \sum_{\mu\lambda} \sum_{\nu \in \mathbf{2}}
\mathrm{Im}\left[ \mathcal{R}^{(a/b)}(\mu,\nu,\lambda;t,\tau) 
\langle 0 | Z_l | \mu \rangle \langle \mu | Z_{l'} | \nu \rangle
\langle \nu | Z_{l'} | \lambda \rangle \langle \lambda | Z_l | 0 \rangle \right] \label{eq:tc_nu2a} \\
&+\frac{1}{L^2} \Theta{(t)}\Theta{(\tau)} \sum_{l \neq l '}\sum_{\mu\lambda}\sum_{\nu \in \mathbf{2}}
\mathrm{Im}\left[ \mathcal{R}^{(a/b)}(\mu,\nu,\lambda;t,\tau) 
\langle 0 | Z_l | \mu \rangle \langle \mu | Z_{l'} | \nu \rangle
\langle \nu | Z_l | \lambda \rangle \langle \lambda | Z_{l'} | 0 \rangle \right]  \label{eq:tc_nu2b} \\
&+\frac{1}{L^2}  \Theta{(t)}\Theta{(\tau)} \sum_{l \neq l'} \sum_{\mu\lambda} \sum_{\nu \in \mathbf{4}}
\mathrm{Im}\left[ \mathcal{R}^{(a/b)}(\mu,\nu,\lambda;t,\tau) 
\langle 0 | Z_l | \mu \rangle \langle \mu | Z_{l'} | \nu \rangle
\langle \nu | Z_{l'} | \lambda \rangle \langle \lambda | Z_l | 0 \rangle \right] \label{eq:tc_nu4a} \\
&+\frac{1}{L^2} \Theta{(t)}\Theta{(\tau)} \sum_{l \neq l'} \sum_{\mu\lambda} \sum_{\nu \in \mathbf{4}}
\mathrm{Im}\left[ \mathcal{R}^{(a/b)}(\mu,\nu,\lambda;t,\tau) 
\langle 0 | Z_l | \mu \rangle \langle \mu | Z_{l'} | \nu \rangle
\langle \nu | Z_l | \lambda \rangle \langle \lambda | Z_{l'} | 0 \rangle \right] \label{eq:tc_nu4b} \\
&+\frac{1}{L^2}  \Theta{(t)}\Theta{(\tau)} \sum_{l_1 l_2 l_3 l_4 \in \msquare}\sum_{\mu\lambda}\sum_{\nu \in \mathbf{2}} \mathrm{Im}\left[ \mathcal{R}^{(a/b)}(\mu,\nu,\lambda;t,\tau) 
\langle 0 | Z_{l_1} | \mu \rangle \langle \mu | Z_{l_2} | \nu \rangle
\langle \nu | Z_{l_3} | \lambda \rangle \langle \lambda | Z_{l_4} | 0 \rangle \right] \label{eq:tc_nu2c} \\
&+\frac{1}{L^2}  \Theta{(t)}\Theta{(\tau)} \sum_{l_1 l_2 l_3 l_4 \in \msquare}\sum_{\mu\lambda} \sum_{\nu\in\mathbf{4}} \mathrm{Im}\left[ \mathcal{R}^{(a/b)}(\mu,\nu,\lambda;t,\tau) 
\langle 0 | Z_{l_1} | \mu \rangle \langle \mu | Z_{l_2} | \nu \rangle
\langle \nu | Z_{l_3} | \lambda \rangle \langle \lambda | Z_{l_4} | 0 \rangle \right], \label{eq:tc_nu4c}
\end{align}
where $\msquare$ denotes the four edges of a plaquette, and $\mathbf{2}$ and $\mathbf{4}$ denote the sets of all two-particle states and four-particle states, respectively.
The last two sums, Eqs.~(\ref{eq:tc_nu2c}) and (\ref{eq:tc_nu4c}), are the consequence of the ground stating being a symmetric superposition of all possible closed loop.

Since a $Z_l$ operator creates or annihilates a pair of two bosonic charges at two ends of the link $l$, $|\mu\rangle$ and $|\lambda\rangle$ must be the two-particle states.
Without any perturbations, the phase factors are simplified to
\begin{align}
\mathcal{R}^{(a)}(\mathbf{2},\nu,\mathbf{2};t,\tau) &= 2 e^{i(\varepsilon_\nu - 2)t}
+ e^{i\varepsilon_\nu\tau}e^{i(\varepsilon_\nu-2)t}+e^{-i \varepsilon_\nu\tau}e^{-2it}, \label{eq:Ra_tc}\\
\mathcal{R}^{(b)}(\mathbf{2},\nu,\mathbf{2};t,\tau) &= e^{-2i\tau}e^{i(\varepsilon_\nu -2)t}
+ 2e^{2i\tau}e^{i(\varepsilon_\nu-2)t} + e^{-2i\tau}e^{-2it},\label{eq:Rb_tc}
\end{align}
where
\begin{equation}
\varepsilon_\nu = E_\nu - E_0 = 
\begin{cases}
0 & \mathrm{if}~|\nu\rangle = |0\rangle \\
2 & \mathrm{if}~|\nu\rangle \in \mathbf{2} \\
4 & \mathrm{if}~|\nu\rangle \in \mathbf{4}. \\
\end{cases}
\end{equation}
Then
\begin{align}
\chi_z^{(a/b)}(t,\tau) &=  \Theta{(t)}\Theta{(\tau)} \mathrm{Im}\left[ (2L^2)\mathcal{R}^{(a/b)}(\mathbf{2},0,\mathbf{2};t,\tau)
+ 6\mathcal{R}^{(a/b)}(\mathbf{2},\mathbf{2},\mathbf{2};t,\tau)
+ 6\mathcal{R}^{(a/b)}(\mathbf{2},\mathbf{2},\mathbf{2};t,\tau) \right. \\
&+(2L^2 - 7)\mathcal{R}^{(a/b)}(\mathbf{2},\mathbf{4},\mathbf{2};t,\tau)
+(2L^2 - 7)\mathcal{R}^{(a/b)}(\mathbf{2},\mathbf{4},\mathbf{2};t,\tau) \\
&\left.+8\mathcal{R}^{(a/b)}(\mathbf{2},\mathbf{2},\mathbf{2};t,\tau) 
+4\mathcal{R}^{(a/b)}(\mathbf{2},\mathbf{4},\mathbf{2};t,\tau)\right] \\
&= (2L^2)  \Theta{(t)}\Theta{(\tau)} \,\mathrm{Im} \left[ \mathcal{R}^{(a/b)}(\mathbf{2},0,\mathbf{2};t,\tau) + 2 \mathcal{R}^{(a/b)}(\mathbf{2},\mathbf{4},\mathbf{2};t,\tau) \right] \label{eq:tc_N2}\\
&+10  \Theta{(t)}\Theta{(\tau)}\left[2 \,\mathrm{Im}\, \mathcal{R}^{(a/b)}(\mathbf{2},\mathbf{2},\mathbf{2};t,\tau)
-\mathrm{Im}\, \mathcal{R}^{(a/b)}(\mathbf{2},\mathbf{4},\mathbf{2};t,\tau) \right]. \label{eq:tc_N}
\end{align}

As we discussed previously, extensive terms in Eq.~(\ref{eq:tc_N2}) need to be cancelled to be consistent with the definition of the nonlinear susceptibility.
One can confirm the cancellation by substituting Eqs.~(\ref{eq:Ra_tc}) and (\ref{eq:Rb_tc}) to Eq.~(\ref{eq:tc_N2}).
Therefore
\begin{align}
\chi_z^{(a)}(t,\tau) &= 40\,\Theta(t)\Theta(\tau)\left[\sin 2\tau - \sin 2t - \sin 2(t+\tau) \right], \label{eq:chi3a_tc} \\
\chi_z^{(b)}(t,\tau) &= 20\, \Theta(t)\Theta(\tau) \left[ 2 \sin 2\tau - \sin 2(t-\tau) -3\sin2(t+\tau) \right].
\label{eq:chi3b_tc}
\end{align}
We can get the expression for $\chi_x^{(a/b)}(t,\tau)$ by redefining the energy cost for a pair of excitations, $2\to 2J$.


In the presence of weak perturbations, $e$ and $m$ particles become dynamical and dispersive.
For simplicity, let's assume that the elementary excitations have the two-dimensional tight-binding dispersion, $E_{e/m}(q_x, q_y)  = \varepsilon_{e/m} + w(\cos q_x + \cos q_y)$, with the full bandwidth $w$ and the single particle energy gap $\varepsilon_e = 1$ and $\varepsilon_m = J$ for $e$ and $m$ particles, respectively.
Since the particles are no longer static, we use the resolution of identity in plane wave basis:
\begin{equation}
\frac{1}{2} \sum_{p_1 p_2}
|p_1,p_2 \rangle\langle p_1,p_2| = 1.
\end{equation}

We focus on $\chi_z^{(a/b)}(t,\tau)$ with $|\nu\rangle \in \mathbf{2}$ because it is the nonlinear response from a pair of deconfined charges without creating additional charges in the middle which is the important process to diagnose the deconfinement, as discussed in the main text:
\begin{multline}
\chi_z^{(a/b)}(\nu \in \mathbf{2}; t,\tau) = \frac{1}{2L^2} \, \Theta(t)\Theta(\tau)
\mathrm{Im}\left[
\frac{1}{2^3} \sum_{p_1 p_2} \sum_{q_1 q_2} \sum_{r_1 r_2}
\mathcal{R}^{(a/b)}(p_1, p_2, q_1, q_2, r_1, r_2; t, \tau) \right. \\
\left.\times \frac{1}{2^4} \sum_{jklm}\sum_{\eta_j\eta_k\eta_l\eta_m}
\langle 0 |Z_{j, j+\eta_j} |p_1, p_2\rangle
\langle p_1, p_2 |Z_{k, k+\eta_k} |q_1, q_2\rangle
\langle q_1, q_2 |Z_{l, l+\eta_l} |r_1, r_2\rangle
\langle r_1, r_2 |Z_{m, m+\eta_m} |0\rangle \right],
\label{eq:chi3nu2}
\end{multline}
where we label the link $l = \langle j, j+\eta_j\rangle$ with the two ends of the link.

Since $Z_{j, j+\eta_j}$ creates or annihilates two hard-core bosons at the two ends $j$ and $j+\eta_j$ ($\eta_j = \pm \hat{x}, \pm \hat{y}$), we represent the $Z$ operator with bosonic creation operators:
\begin{align}
Z_{j,j+\eta_j}| \, \cdot \, \rangle &\to
a^\dagger_j a^\dagger_{j+\eta_j} | \, \cdot \, \rangle.
\end{align}
However, above representation needs to be carefully modified when $Z_{j,j+\eta_j}$ is acting on a state which already have a boson at either $j$ or $j+\eta_j$. 
To properly represent the spin operator in terms of bosons, we introduce the constraint
\begin{equation}
(a_j^\dagger)^2 \to a_j a_j^\dagger = 1
\end{equation}
because $Z_{j,j+\eta_j}$ operator annihilate the boson at site $j$ if the site is already occupied.
Therefore
\begin{align}
\sum_{j,\eta_j} \langle \xi_1, \xi_2 | Z_{j,j+\eta_j} | 0 \rangle
&\to \sum_{j,\eta_j} \langle \xi_1, \xi_2 | a_j^\dagger a_{j+\eta_j}^\dagger |0\rangle
= \sum_{j,\eta_j} \langle 0 | a_{\xi_1} a_{\xi_2} a_j^\dagger a_{j+\eta_j}^\dagger |0\rangle
= \sum_{j,\eta_j} \delta_{\xi_1,j}\delta_{\xi_2,j+\eta_j} + \delta_{\xi_1,j+\eta_j}\delta_{\xi_2,j} \\
&=\sum_{j,\eta_j} 2\delta_{\xi_1,j}\delta_{\xi_2,j+\eta_j}, \label{eq:0to2}
\end{align}
and
\begin{align}
\sum_{k,\eta_k} \langle \xi_1, \xi_2 | Z_{k,k+\eta_k} |\zeta_1, \zeta_2\rangle
&\to \sum_{k,\eta_k} \langle 0 | a_{\xi_1} a_{\xi_2} a_k^\dagger a_{k+\eta_k}^\dagger a_{\zeta_1}^\dagger a_{\zeta_2}^\dagger |0\rangle \\
&\to \sum_{k,\eta_k} \delta_{\zeta_1, k}\langle a_{\xi_1} a_{\xi_2} a_{k+\eta_k}^\dagger a_{\zeta_2}^\dagger\rangle
+ \delta_{\zeta_2, k}\langle a_{\xi_1} a_{\xi_2} a_{k+\eta_k}^\dagger a_{\zeta_1}^\dagger\rangle 
+ \delta_{\zeta_1, k+\eta_k}\langle a_{\xi_1} a_{\xi_2} a_{k}^\dagger a_{\zeta_2}^\dagger\rangle
\nonumber \\
&\qquad
+ \delta_{\zeta_2, k+\eta_k}\langle a_{\xi_1} a_{\xi_2} a_{k}^\dagger a_{\zeta_1}^\dagger\rangle \\
&= \sum_{k,\eta_k}\delta_{\zeta_1, k}(\delta_{\xi_1, k+\eta_k} \delta_{\xi_2,\zeta_2} + \delta_{\xi_1, \zeta_2}\delta_{\xi_2,k+\eta_k})
+ \delta_{\zeta_2, k}(\delta_{\xi_1, k+\eta_k} \delta_{\xi_2,\zeta_1} + \delta_{\xi_1, \zeta_1}\delta_{\xi_2,k+\eta_k}) \nonumber \\
&\qquad+\delta_{\zeta_1, k+\eta_k}(\delta_{\xi_1, k} \delta_{\xi_2,\zeta_2} + \delta_{\xi_1, \zeta_2}\delta_{\xi_2,k})
+ \delta_{\zeta_2, k+\eta_k}(\delta_{\xi_1, k} \delta_{\xi_2,\zeta_1} + \delta_{\xi_1, \zeta_1}\delta_{\xi_2,k}) \\
&=\sum_{k,\eta_k} 2 \left[
\delta_{\zeta_1, k}(\delta_{\xi_1, k+\eta_k} \delta_{\xi_2,\zeta_2} + \delta_{\xi_1, \zeta_2}\delta_{\xi_2,k+\eta_k})
+ \delta_{\zeta_2, k}(\delta_{\xi_1, k+\eta_k} \delta_{\xi_2,\zeta_1} + \delta_{\xi_1, \zeta_1}\delta_{\xi_2,k+\eta_k})
\right]. \label{eq:2to2}
\end{align}

Using Eqs.~(\ref{eq:0to2}) and (\ref{eq:2to2}), we can calculate the matrix elements to calculate the susceptibility in Eq.~(\ref{eq:chi3nu2})
\begin{align}
\frac{1}{2} \sum_{j,\eta_j} \langle 0 |Z_{j,j+\eta_j} |p_1,p_2\rangle &=
\frac{1}{2L^2} \sum_{j, \eta_j} \sum_{\xi_1 \neq \xi_2}
e^{i p_1 \cdot \xi_1 + i p_2 \cdot \xi_2}  \langle 0| Z_{j,j+\eta_j} | \xi_1,\xi_2\rangle \\
&= \frac{1}{L^2} \sum_{j, \eta_j} \sum_{\xi_1 \neq \xi_2}
e^{i p_1 \cdot \xi_1 + i p_2 \cdot \xi_2} \delta_{\xi_1,j}\delta_{\xi_2,j+\eta_j}\\
&= \frac{1}{L^2} \sum_{j,\eta_j}
e^{i(p_1+p_2)\cdot R_j} e^{i p_2 \cdot \eta_j}
= 2(\cos p_{1x} + \cos p_{1y}) \delta_{p_1 + p_2, 0},
\label{eq:0Zp}
\end{align}
and
\begin{align}
\frac{1}{2}\sum_{k,\eta_k} \langle p_1,p_2|Z_{k,k+\eta_k} |q_1, q_2\rangle &=
\frac{1}{2} \left(\frac{1}{L^2}\right)^2 \sum_{k,k+\eta_k} \sum_{\xi_1 \neq \xi_2}
\sideset{}{'}\sum_{\zeta_1, \zeta_2} e^{-ip_1\cdot \xi_1 - i p_2 \cdot \xi_2}e^{iq_1 \cdot \zeta_1 + iq_2 \cdot \zeta_2}
\langle \xi_1, \xi_2 | Z_{k,k+\eta_k} |\zeta_1, \zeta_2 \rangle \\
&= \left( \frac{1}{L^2} \right)^2 \sum_{k,\eta_k} \sum_{\xi_1 \neq \xi_2}
\sideset{}{'}\sum_{\zeta_1, \zeta_2}  e^{-ip_1\cdot \xi_1 - i p_2 \cdot \xi_2}
e^{iq_1 \cdot \zeta_1 + iq_2 \cdot \zeta_2}
\left[
\delta_{\zeta_1, k}(\delta_{\xi_1, k+\eta_k} \delta_{\xi_2,\zeta_2} + \delta_{\xi_1, \zeta_2}\delta_{\xi_2,k+\eta_k}) \right. \nonumber \\
&\qquad\qquad\qquad\qquad\qquad\qquad\qquad\qquad\qquad
\left.+ \delta_{\zeta_2, k}(\delta_{\xi_1, k+\eta_k} \delta_{\xi_2,\zeta_1} + \delta_{\xi_1, \zeta_1}\delta_{\xi_2,k+\eta_k})
\right] \\
&=\frac{1}{L^4} \sum_{k,\eta_k} \sum_{\zeta_1\neq k,k+\eta_k}
e^{i(q_2 - p_1) \cdot R_k} e^{-ip_1 \cdot \eta_k} e^{i(q_1 - p_2)\cdot \zeta_1}
+ e^{i(q_2 - p_2) \cdot R_k} e^{-ip_2 \cdot \eta_k} e^{i(q_1 - p_1)\cdot \zeta_1} \nonumber \\
&+\frac{1}{L^4} \sum_{k,\eta_k} \sum_{\zeta_2\neq k,k+\eta_k}
e^{i(q_1 - p_1)\cdot R_k}e^{-ip_1 \cdot \eta_k} e^{i(q_2 - p_2)\cdot \zeta_2}
+ e^{i(q_1 - p_2)\cdot R_k} e^{-ip_2 \cdot \eta_k} e^{i(q_2 - p_1)\cdot \zeta_2} \\
&= \delta_{q_2 p_1} \delta_{q_1 p_2} e^{-ip_1 \cdot \eta_k}
+ \delta_{q_2 p_2} \delta_{q_1 p_1} e^{-ip_2 \cdot \eta_k}
+\delta_{q_1 p_1} \delta_{q_2 p_2} e^{-ip_1 \cdot \eta_k}
+\delta_{q_1 p_2} \delta_{q_2 p_1} e^{-ip_2 \cdot \eta_k} \\
&- \frac{1}{L^4} \sum_{k,\eta_k} e^{i(q_1 + q_2 - p_1 - p_2) \cdot R_k} \left[
e^{-ip_1 \cdot \eta_k} \left( 2 + e^{i(q_1- p_2)\cdot \eta_k} \right)
+e^{-ip_2 \cdot \eta_k} \left( 2 + e^{i(q_1- p_1)\cdot \eta_k} \right) \right. \nonumber\\
&\qquad\qquad\qquad\qquad\qquad
\left.+e^{-ip_1 \cdot \eta_k} \left( 2 + e^{i(q_2- p_2)\cdot \eta_k} \right)
+e^{-ip_2 \cdot \eta_k} \left( 2 + e^{i(q_2- p_1)\cdot \eta_k} \right)
\right] \\
&=
2(\delta_{q_1 p_1}\delta_{q_2 p_2} + \delta_{q_1 p_2}\delta_{q_2 p_1})
(\cos p_{1x} + \cos p_{1y} + \cos p_{2x} + \cos p_{2y}) \label{eq:tc_noscattering}\\
&- \frac{4}{L^2} \delta_{q_1+q_2, p_1 + p_2}
\left(
\cos p_{1x} + \cos p_{1y} + \cos p_{2x} + \cos p_{2y} \right.\nonumber \\
&\qquad\qquad\qquad\qquad\qquad\qquad\qquad\qquad\qquad
\left.+\cos q_{1x} + \cos q_{1y} + \cos q_{2x} + \cos q_{2y}
 \right).\label{eq:tc_scattering}
\end{align}
Eq.~(\ref{eq:tc_noscattering}) represents freely propagating process of two hard core bosons while
Eq.~(\ref{eq:tc_scattering}) illustrates the elastic scattering of two hard core bosons without creating additional particles.

Since the Kronecker deltas are imposing the conservation of total momentum,
let's denote $p \equiv p_1$, $p_2 = -p$, $q\equiv q_1$, and $q_2 = -q$.
Also note that the phase factor $\mathcal{R}(p,-p,q,-q,r,-r;t,\tau)$ does not depend on the overall sign of momenta, \textit{i.e.}, $\mathcal{R}(p,-p,q,-q,r,-r;t,\tau) = \mathcal{R}(-p,p,q,-q,r,-r;t,\tau) = ... =\mathcal{R}(-p,p,-q,q,-r,r;t,\tau)$.
Hence, for concise notation, let's write $\mathcal{R}(p,q,r;t,\tau) \equiv \mathcal{R}(p,-p,q,-q,r,-r;t,\tau)$.
Then
\begin{align}
\chi_z^{(a/b)}(\nu\in\mathbf{2};t,\tau) &=
\frac{16}{L^2}\sum_p \mathcal{R}(p,p,p;t,\tau) \left( \cos p_x + \cos p_y \right)^4 \nonumber \\
&-\frac{16}{L^4} \sum_{p,r}\mathcal{R}(p,p,r;t,\tau) \left(\cos p_x + \cos p_y \right)^2
\left( \cos p_x + \cos p_y + \cos r_x + \cos r_y \right) \left(\cos r_x + \cos r_y\right) \nonumber \\
&-\frac{16}{L^4} \sum_{p,r}\mathcal{R}(p,r,r;t,\tau) \left(\cos p_x + \cos p_y\right)
\left( \cos p_x + \cos p_y + \cos r_x + \cos r_y\right) \left(\cos r_x + \cos r_y\right)^2 \nonumber \\
&+\frac{16}{L^6} \sum_{p,q,r} \mathcal{R}(p,q,r;t,\tau) \left(\cos p_x + \cos p_y\right)
\left( \cos p_x + \cos p_y + \cos q_x + \cos q_y \right) \nonumber\\
&\qquad\qquad\qquad\qquad\qquad\qquad\qquad\qquad \times
\left( \cos q_x + \cos q_y + \cos r_x + \cos r_y \right)
\left(\cos r_x + \cos r_y\right),
\label{eq:chi3nu2_2d}
\end{align}
with
\begin{align}
\mathcal{R}^{(a)}(p,q,r;t,\tau) &= 2 e^{2i(\varepsilon_p-\varepsilon_r)\tau}e^{2i(\varepsilon_q - \varepsilon_r)t}
+ e^{2i \varepsilon_q\tau}e^{2i(\varepsilon_q-\varepsilon_r)t}+e^{-2i\varepsilon_q\tau}e^{-2i\varepsilon_pt}, \label{eq:Ra_p}\\
\mathcal{R}^{(b)}(p,q,r;t,\tau) &= e^{-2i \varepsilon_r\tau}e^{2i(\varepsilon_q - \varepsilon_r)t}
+ 2e^{2i\varepsilon_p\tau}e^{2i(\varepsilon_q-\varepsilon_r)t} + e^{-2i\varepsilon_r\tau}e^{-2i\varepsilon_p t},\label{eq:Rb_p}
\end{align}
where $\varepsilon_p = 1 + w(\cos p_x + \cos p_y)$.
If $w = 0$, Eq.~(\ref{eq:chi3nu2_2d}) reproduces the third order susceptibilities of the unperturbed toric code for $\nu \in \mathbf{2}$:
\begin{align}
\chi^{(a)}_z(\nu\in\mathbf{2};t,\tau) = \chi^{(b)}_z(\nu\in\mathbf{2};t,\tau)
= 40 \,\Theta(t)\Theta(\tau) \left[ \sin 2\tau - \sin 2(t+\tau) \right].
\end{align}

\subsection{X-cube model}

X-cube model is one of the prototypical lattice models for type-I fracton system.
The magnetic fields along $x$-axis excite a pair of lineons $l_\alpha$ at two vertices, whose motion is constrained to one dimensional line along $\alpha$-axis.
The fields along $z$-axis excite cubes called fractons which cannot move unless two fractons pair up to a composite object called planons $P_{\alpha\beta}$. Planons $P_{\alpha\beta}$ also has restricted mobility within a two-dimensional $\alpha\beta$ plane.

\subsubsection{Nonlinear dynamics of lineons}

The third-order susceptibilities $\chi_{xxxx}(t,t+\tau,t+\tau) \equiv \chi^{(a)}_x(t,\tau)$ and $\chi_{xxxx}(t,t,t+\tau) \equiv \chi^{(b)}_x(t,\tau)$ are quantifying nonlinear response of the lineons.
Without any perturbations, the susceptibilities can be exactly calculated with the resolution of identity in the real space basis:
\begin{align}
\chi_x^{(a/b)}(t,\tau) &= \frac{2}{3L^3} \Theta{(t)}\Theta{(\tau)} \sum_{l,l'} \sum_{\mu\lambda}
\mathrm{Im}\left[ \mathcal{R}^{(a/b)}(\mu,0,\lambda;t,\tau) 
\langle 0 | X_l | \mu \rangle \langle \mu | X_l | 0 \rangle
\langle 0 | X_{l'} | \lambda \rangle \langle \lambda | X_{l'} | 0 \rangle \right] \label{eq:lineon_nu0} \\
&+ \frac{2}{3L^3}  \Theta{(t)}\Theta{(\tau)} \sum_{l \neq l'} \sum_{\mu\lambda} \sum_{\nu \in \mathbf{2}}
\mathrm{Im}\left[ \mathcal{R}^{(a/b)}(\mu,\nu,\lambda;t,\tau) 
\langle 0 | X_l | \mu \rangle \langle \mu | X_{l'} | \nu \rangle
\langle \nu | X_{l'} | \lambda \rangle \langle \lambda | X_l | 0 \rangle \right] \label{eq:lineon_nu2a} \\
&+\frac{2}{3L^3} \Theta{(t)}\Theta{(\tau)} \sum_{l \neq l '}\sum_{\mu\lambda}\sum_{\nu \in \mathbf{2}}
\mathrm{Im}\left[ \mathcal{R}^{(a/b)}(\mu,\nu,\lambda;t,\tau) 
\langle 0 | X_l | \mu \rangle \langle \mu | X_{l'} | \nu \rangle
\langle \nu | X_l | \lambda \rangle \langle \lambda | X_{l'} | 0 \rangle \right]  \label{eq:lineon_nu2b} \\
&+\frac{2}{3L^3}  \Theta{(t)}\Theta{(\tau)} \sum_{l \neq l'} \sum_{\mu\lambda} \sum_{\nu \in \mathbf{4}}
\mathrm{Im}\left[ \mathcal{R}^{(a/b)}(\mu,\nu,\lambda;t,\tau) 
\langle 0 | X_l | \mu \rangle \langle \mu | X_{l'} | \nu \rangle
\langle \nu | X_{l'} | \lambda \rangle \langle \lambda | X_l | 0 \rangle \right] \label{eq:lineon_nu4a} \\
&+\frac{2}{3L^3} \Theta{(t)}\Theta{(\tau)} \sum_{l \neq l'} \sum_{\mu\lambda} \sum_{\nu \in \mathbf{4}}
\mathrm{Im}\left[ \mathcal{R}^{(a/b)}(\mu,\nu,\lambda;t,\tau) 
\langle 0 | X_l | \mu \rangle \langle \mu | X_{l'} | \nu \rangle
\langle \nu | X_l | \lambda \rangle \langle \lambda | X_{l'} | 0 \rangle \right] \label{eq:lineon_nu4b} \\
&=6L^3 \Theta{(t)}\Theta{(\tau)} \mathrm{Im}\, \mathcal{R}^{(a/b)}(\mathbf{2},0,\mathbf{2};t,\tau) \\
&+4 \Theta{(t)}\Theta{(\tau)} \mathrm{Im}\,\mathcal{R}^{(a/b)}(\mathbf{2},\mathbf{2},\mathbf{2};t,\tau)
+4 \Theta{(t)}\Theta{(\tau)} \mathrm{Im}\,\mathcal{R}^{(a/b)}(\mathbf{2},\mathbf{2},\mathbf{2};t,\tau) \\
&+2(3L^3-3) \Theta{(t)}\Theta{(\tau)} \mathrm{Im}\,\mathcal{R}^{(a/b)}(\mathbf{2},\mathbf{4},\mathbf{2};t,\tau)
+2(3L^3-3) \Theta{(t)}\Theta{(\tau)} \mathrm{Im}\,\mathcal{R}^{(a/b)}(\mathbf{2},\mathbf{4},\mathbf{2};t,\tau) \\
&=6L^3 \Theta(t)\Theta(\tau)\mathrm{Im}\left[\mathcal{R}^{(a/b)}(\mathbf{2},0,\mathbf{2};t,\tau) + 2 \mathcal{R}^{(a/b)}(\mathbf{2},\mathbf{4},\mathbf{2};t,\tau)\right] \label{eq:lineon_L3} \\
&+ 4 \Theta(t)\Theta(\tau) \mathrm{Im} \left[ 2\mathcal{R}^{(a/b)}(\mathbf{2},\mathbf{2},\mathbf{2};t,\tau) - 3\mathcal{R}^{(a/b)}(\mathbf{2},\mathbf{4},\mathbf{2};t,\tau) \right].
\end{align}
Like 2D toric code, the terms proportional to the system size $L^3$ are vanishing [Eq.~(\ref{eq:lineon_L3})$\, =0$].
With a single lineon excitation energy $\varepsilon_l = 2$, the third-order susceptibilities are
\begin{align}
\chi_x^{(a)}(t,\tau) &= 8 \theta(t)\theta(\tau) \left[ \sin 4\tau - \sin 4(t+\tau) - 3 \sin 4t  \right], \label{eq:lineon1} \\
\chi_x^{(b)}(t,\tau) &= 4 \theta(t)\theta(\tau) \left[ 2\sin 4\tau - 5\sin 4(t+\tau) - 3 \sin 4(t-\tau)  \right].\label{eq:lineon2}
\end{align}

In the presence of weak perturbations, the nonlinear susceptibilities can be derived by following the similar calculations done for the toric code. Again we focus on the case where $|\nu\rangle \in \mathbf{2}$:
\begin{align}
\chi^{(a/b)}(\nu \in \mathbf{2};t,\tau) &=
\frac{64}{L} \sum_p \mathcal{R}^{(a/b)}(p,p,p;t,\tau) \cos^4 p \\
&-\frac{64}{L^2} \sum_{p,r} \left[\mathcal{R}^{(a/b)}(p,p,r;t,\tau)\cos^2 p (\cos p + \cos r) \cos r
+\mathcal{R}^{(a/b)}(p,r,r;t,\tau) \cos p (\cos p + \cos r) \cos^2 r \right] \\
&+\frac{64}{L^3} \sum_{p,q,r} \mathcal{R}^{(a/b)}(p,q,r;t,\tau) \cos p (\cos p + \cos q) (\cos q + \cos r) \cos r,
\end{align}
where $\mathcal{R}^{(a/b)}(p,q,r;t,\tau)$ are calculated from Eqs.~(\ref{eq:Ra_p}) and (\ref{eq:Rb_p}) with the one-dimensional dispersion for lineons, $\varepsilon_l(p) = 2 + w \cos p$.
This reproduces $\nu \in \mathbf{2}$ contributions of Eq.(\ref{eq:lineon1}) and (\ref{eq:lineon2}) when $w = 0$. 

\subsubsection{Nonlinear dynamics of planons and fractons}

\begin{figure}[h]
\centering
\includegraphics[width=0.8\textwidth]{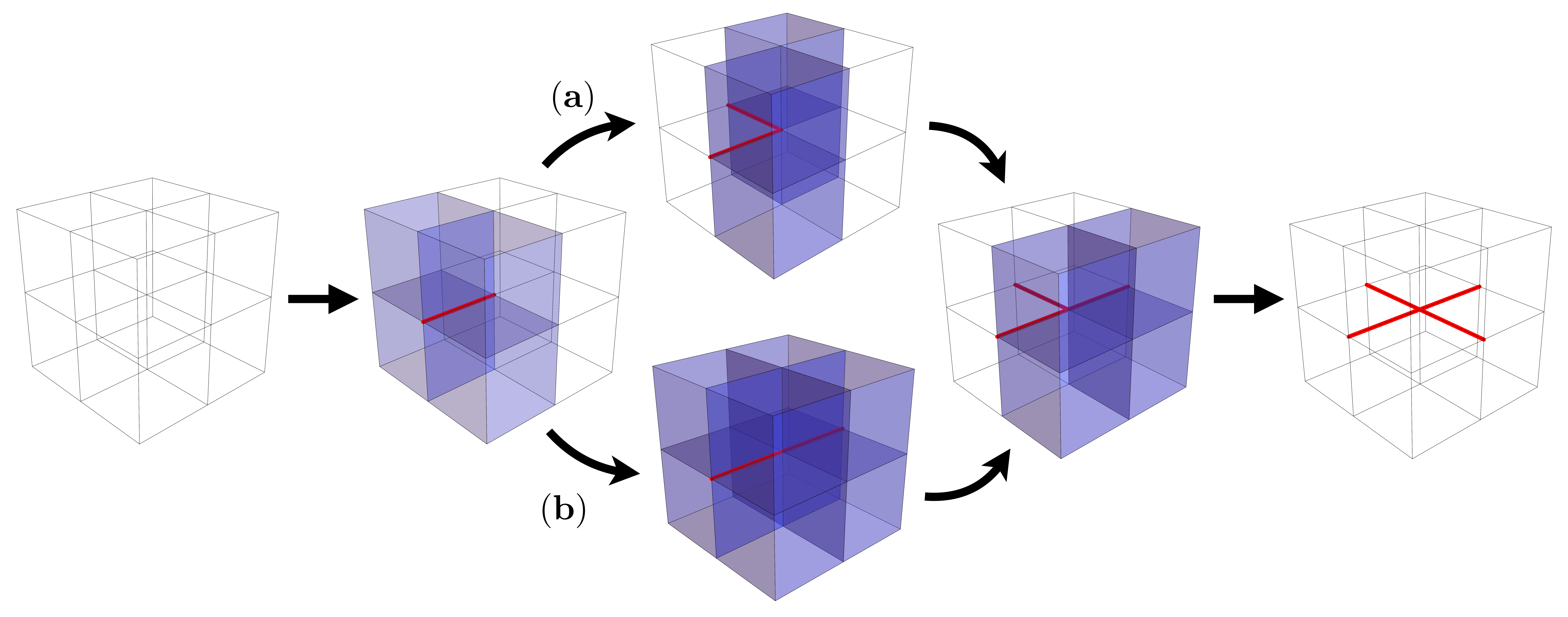}
\caption{The ground state of X-cube model is the symmetric sum of all possible spin configurations having zero excitation. So there are four-spin-flip processes which connect two inequivalent vacuum spin configurations. In the middle of the perturbative process, (a) only four cube excited states can be involved, or (b) an eight cube excited state can be involved.}
\label{fig:cross}
\end{figure}

An action of a single $Z$ operator on the ground state excites four fractons which can be grouped into a pair of two planons [FIG. \ref{fig:cross} (a)].
Nonlinear dynamics of the planons and immobile fractons are characterized with the third-order susceptibilities $\chi_{zzzz}(t,t+\tau,t+\tau) \equiv \chi^{(a)}_z(t,\tau)$ and $\chi_{zzzz}(t,t,t+\tau) \equiv \chi^{(b)}_z(t,\tau)$.

\begin{figure}[h]
\centering
\includegraphics[width=\linewidth]{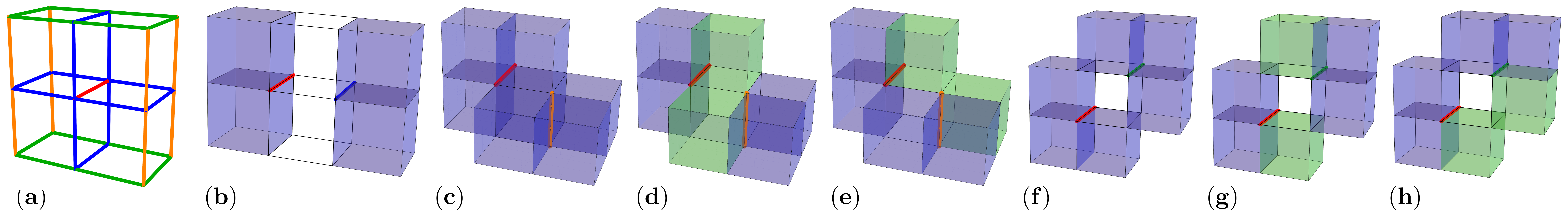}
\caption{(a) After $Z$ acts on the red link,  different fracton configurations are available depending on which link is flipped subsequently. (b) If a blue link $l \in \mathcal{B}$ is flipped, a planon (face-sharing two-cube composite) is displaced by a unit distance.
When (c) an orange link $l \in \mathcal{O}$ or (f) a green link $l \in \mathcal{G}$ is flipped, six fractons are excited.
(d, e, g, h) These fractons can be grouped into two mobile planons (blue face-sharing cubes) and two immobile fractons (green cubes). While the two planons in (g) have restricted mobility within two parallel planes, the planons in (d, e, h) are constrained to two planes perpendicular to each other.
If the secondly flipped link does not belong to the set of nearby links $\mathcal{B} \cup \mathcal{O}\cup \mathcal{G}$, then total eight fractons are excited by the two spin flips.}
\label{fig:fractons}
\end{figure}

In the absence of perturbations, we again calculate the exact nonlinear susceptibilities with the real space basis:
\begin{align}
\chi_z^{(a/b)}(t,\tau) &= \frac{2}{3L^3} \Theta{(t)}\Theta{(\tau)} \sum_{l,l'} \sum_{\mu\lambda}
\mathrm{Im}\left[ \mathcal{R}^{(a/b)}(\mu,0,\lambda;t,\tau) 
\langle 0 | Z_l | \mu \rangle \langle \mu | Z_l | 0 \rangle
\langle 0 | Z_{l'} | \lambda \rangle \langle \lambda | Z_{l'} | 0 \rangle \right] \\
&+ \frac{2}{3L^3}   \Theta{(t)}\Theta{(\tau)} \sum_{l} \sum_{l' \in \mathcal{B}} \sum_{\mu\nu\lambda}
\mathrm{Im}\left[ \mathcal{R}^{(a/b)}(\mu,\nu,\lambda;t,\tau) 
\langle 0 | Z_l | \mu \rangle \langle \mu | Z_{l'} | \nu \rangle
\langle \nu | Z_{l'} | \lambda \rangle \langle \lambda | Z_l | 0 \rangle \right] \label{eq:Ba} \\
&+\frac{2}{3L^3}  \Theta{(t)}\Theta{(\tau)} \sum_{l} \sum_{l' \in \mathcal{B}} \sum_{\mu\nu\lambda}
\mathrm{Im}\left[ \mathcal{R}^{(a/b)}(\mu,\nu,\lambda;t,\tau) 
\langle 0 | Z_l | \mu \rangle \langle \mu | Z_{l'} | \nu \rangle
\langle \nu | Z_l | \lambda \rangle \langle \lambda | Z_{l'} | 0 \rangle \right] \label{eq:Bb} \\
&+\frac{2}{3L^3}   \Theta{(t)}\Theta{(\tau)}\sum_{l} \sum_{l' \in \mathcal{G}} \sum_{\mu\nu\lambda}
\mathrm{Im}\left[ \mathcal{R}^{(a/b)}(\mu,\nu,\lambda;t,\tau) 
\langle 0 | Z_l | \mu \rangle \langle \mu | Z_{l'} | \nu \rangle
\langle \nu | Z_{l'} | \lambda \rangle \langle \lambda | Z_l | 0 \rangle \right] \label{eq:Ga} \\
&+\frac{2}{3L^3}  \Theta{(t)}\Theta{(\tau)} \sum_{l} \sum_{l' \in \mathcal{G}} \sum_{\mu\nu\lambda}
\mathrm{Im}\left[ \mathcal{R}^{(a/b)}(\mu,\nu,\lambda;t,\tau) 
\langle 0 | Z_l | \mu \rangle \langle \mu | Z_{l'} | \nu \rangle
\langle \nu | Z_l | \lambda \rangle \langle \lambda | Z_{l'} | 0 \rangle \right] \label{eq:Gb} \\
&+\frac{2}{3L^3}   \Theta{(t)}\Theta{(\tau)}\sum_{l} \sum_{l' \in \mathcal{O}} \sum_{\mu\nu\lambda}
\mathrm{Im}\left[ \mathcal{R}^{(a/b)}(\mu,\nu,\lambda;t,\tau) 
\langle 0 | Z_l | \mu \rangle \langle \mu | Z_{l'} | \nu \rangle
\langle \nu | Z_{l'} | \lambda \rangle \langle \lambda | Z_l | 0 \rangle \right] \label{eq:Oa} \\
&+\frac{2}{3L^3}  \Theta{(t)}\Theta{(\tau)} \sum_{l} \sum_{l' \in \mathcal{O}} \sum_{\mu\nu\lambda}
\mathrm{Im}\left[ \mathcal{R}^{(a/b)}(\mu,\nu,\lambda;t,\tau) 
\langle 0 | Z_l | \mu \rangle \langle \mu | Z_{l'} | \nu \rangle
\langle \nu | Z_l | \lambda \rangle \langle \lambda | Z_{l'} | 0 \rangle \right] \label{eq:Ob} \\
&+\frac{2}{3L^3}   \Theta{(t)}\Theta{(\tau)}\sum_{l} \sum_{l' \neq l, l'\notin \mathcal{B}\cup\mathcal{O}\cup\mathcal{G}} \sum_{\mu\nu\lambda}
\mathrm{Im}\left[ \mathcal{R}^{(a/b)}(\mu,\nu,\lambda;t,\tau) 
\langle 0 | Z_l | \mu \rangle \langle \mu | Z_{l'} | \nu \rangle
\langle \nu | Z_{l'} | \lambda \rangle \langle \lambda | Z_l | 0 \rangle \right] \label{eq:xc_8a}\\
&+\frac{2}{3L^3}  \Theta{(t)}\Theta{(\tau)} \sum_{l} \sum_{l' \neq l, l' \notin \mathcal{B}\cup\mathcal{O}\cup\mathcal{G}} \sum_{\mu\nu\lambda}
\mathrm{Im}\left[ \mathcal{R}^{(a/b)}(\mu,\nu,\lambda;t,\tau) 
\langle 0 | Z_l | \mu \rangle \langle \mu | Z_{l'} | \nu \rangle
\langle \nu | Z_l | \lambda \rangle \langle \lambda | Z_{l'} | 0 \rangle \right] \label{eq:xc_8b}\\
&+\frac{2}{3L^3}   \Theta{(t)}\Theta{(\tau)} \sum_{l_1 l_2 l_3 l_4 \in +}\sum_{\mu\lambda}\sum_{\nu \in \mathbf{4}}
\mathrm{Im}\left[ \mathcal{R}^{(a/b)}(\mu,\nu,\lambda;t,\tau) 
\langle 0 | Z_{l_1} | \mu \rangle \langle \mu | Z_{l_2} | \nu \rangle
\langle \nu | Z_{l_3} | \lambda \rangle \langle \lambda | Z_{l_4} | 0 \rangle \right] \label{eq:xc_cross4} \\
&+\frac{2}{3L^3}   \Theta{(t)}\Theta{(\tau)} \sum_{l_1 l_2 l_3 l_4 \in +}\sum_{\mu\lambda} \sum_{\nu\in\mathbf{8}}
\mathrm{Im}\left[ \mathcal{R}^{(a/b)}(\mu,\nu,\lambda;t,\tau) 
\langle 0 | Z_{l_1} | \mu \rangle \langle \mu | Z_{l_2} | \nu \rangle
\langle \nu | Z_{l_3} | \lambda \rangle \langle \lambda | Z_{l_4} | 0 \rangle \right], \label{eq:xc_cross8}
\end{align}
where the sets $\mathcal{B}$, $\mathcal{O}$, $\mathcal{G}$ are indicated in FIG.~\ref{fig:fractons} (a), and $\mathbf{n}$ is a set of all $n$-cube states.
The last two terms, Eqs.~(\ref{eq:xc_cross4}) and (\ref{eq:xc_cross8}), are consequences of the ground state of X-cube model being a symmetric sum of all inequivalent spin configurations without any excitation (FIG.~\ref{fig:cross}).

Since $Z$ operator excites four cubes, $|\mu\rangle$ and $|\lambda\rangle$ must be the four-cube states. The state $|\nu\rangle$ depends on the link $l'$:
\begin{align}
|\nu\rangle \in
\begin{cases}
\mathbf{4}, & l' \in \mathcal{B}, \\
\mathbf{6}, & l' \in \mathcal{O} \cup \mathcal{G}, \\
\mathbf{8}, & l' \neq l, l' \notin \mathcal{B} \cup \mathcal{O} \cup \mathcal{G}.
\end{cases}
\end{align}
Hence,
\begin{align}
\chi^{(a/b)}_z(t,\tau) &= 2
\Theta(t)\Theta(\tau) \left[ (3L^3)\, \mathrm{Im}\,\mathcal{R}^{(a/b)}(\mathbf{4},0,\mathbf{4};t,\tau)\right] \\
&+2\Theta(t)\Theta(\tau) \left[ 12 \, \mathrm{Im}\,\mathcal{R}^{(a/b)}(\mathbf{4},\mathbf{4},\mathbf{4};t,\tau) \right]  \\
&+2\Theta(t)\Theta(\tau) \left[ 12 \, \mathrm{Im}\,\mathcal{R}^{(a/b)}(\mathbf{4},\mathbf{4},\mathbf{4};t,\tau) \right]  \\
&+2\Theta(t)\Theta(\tau) \left[ 12 \, \mathrm{Im}\,\mathcal{R}^{(a/b)}(\mathbf{4},\mathbf{6},\mathbf{4};t,\tau) \right]  \\
&+2\Theta(t)\Theta(\tau) \left[ 12 \, \mathrm{Im}\,\mathcal{R}^{(a/b)}(\mathbf{4},\mathbf{6},\mathbf{4};t,\tau) \right]  \\
&+2\Theta(t)\Theta(\tau) \left[ 8 \, \mathrm{Im}\,\mathcal{R}^{(a/b)}(\mathbf{4},\mathbf{6},\mathbf{4};t,\tau) \right]  \\
&+2\Theta(t)\Theta(\tau) \left[ 8 \, \mathrm{Im}\,\mathcal{R}^{(a/b)}(\mathbf{4},\mathbf{6},\mathbf{4};t,\tau)\right]  \\
&+2\Theta(t)\Theta(\tau) \left[ (3L^3 - 33) \, \mathrm{Im}\,\mathcal{R}^{(a/b)}(\mathbf{4},\mathbf{8},\mathbf{4};t,\tau) \right]  \\
&+2\Theta(t)\Theta(\tau) \left[ (3L^3 - 33) \, \mathrm{Im}\,\mathcal{R}^{(a/b)}(\mathbf{4},\mathbf{8},\mathbf{4};t,\tau) \right]  \\
&+2\Theta(t)\Theta(\tau) \left[ 16 \, \mathrm{Im}\,\mathcal{R}^{(a/b)}(\mathbf{4},\mathbf{4},\mathbf{4};t,\tau) \right] \\
&+2\Theta(t)\Theta(\tau) \left[  8\, \mathrm{Im}\,\mathcal{R}^{(a/b)}(\mathbf{4},\mathbf{8},\mathbf{4};t,\tau) \right] \\
&= 6L^3 \,\Theta(t)\Theta(\tau) \,\mathrm{Im}\left[ \mathcal{R}^{(a/b)}(\mathbf{4},0,\mathbf{4};t,\tau) + 2 \mathcal{R}^{(a/b)}(\mathbf{4},\mathbf{8},\mathbf{4};t,\tau) \right] \label{eq:xc_extensive} \\
&+ 80 \, \mathrm{Im}\,\mathcal{R}^{(a/b)}(\mathbf{4},\mathbf{4},\mathbf{4};t,\tau)
+ 80 \, \mathrm{Im}\,\mathcal{R}^{(a/b)}(\mathbf{4},\mathbf{6},\mathbf{4};t,\tau)
-116\, \mathrm{Im}\,\mathcal{R}^{(a/b)}(\mathbf{4},\mathbf{8},\mathbf{4};t,\tau),\label{eq:chi_xcube_z}
\end{align}
where
\begin{align}
\mathcal{R}^{(a)}(\mathbf{4},\mathbf{n},\mathbf{4};t,\tau) &= 2 e^{i(n - 4)Kt}
+ e^{inK\tau}e^{i(n-4)Kt}+e^{-i n K\tau}e^{-4iKt}, \label{eq:Ra_xc}\\
\mathcal{R}^{(b)}(\mathbf{4},\mathbf{n},\mathbf{4};t,\tau) &= e^{-4iK\tau}e^{i(n -4)Kt}
+ 2e^{4iK\tau}e^{i(n-4)Kt} + e^{-4iK\tau}e^{-4iKt}. \label{eq:Rb_xc}
\end{align}
We can check that the extensive term, Eq.~(\ref{eq:xc_extensive}), is vanishing and only the system size independent terms survive.
Therefore
\begin{align}
\chi_z^{(a)}(t,\tau) &=
\chi_z^{(a)}(\nu\in\mathbf{4}; t,\tau) + \chi_z^{(a)}(\nu\in\mathbf{6}; t,\tau) + \chi_z^{(a)}(\nu\in\mathbf{8}; t,\tau) \\
&=80\, \Theta(t)\Theta(\tau) \left[ \sin 4K\tau - \sin 4K(t+\tau) \right] \\
&+ 80\, \Theta(t)\Theta(\tau) \left[ 2 \sin 2Kt + \sin K(2t + 6\tau) - \sin K(4t + 6\tau)  \right] \\
&- 232\, \Theta(t)\Theta(\tau) \sin 4Kt, \\
\chi_z^{(b)}(t,\tau) &=
\chi_z^{(b)}(\nu\in\mathbf{4}; t,\tau) + \chi_z^{(b)}(\nu\in\mathbf{6}; t,\tau) + \chi_z^{(b)}(\nu\in\mathbf{8}; t,\tau) \\
&=
80 \, \Theta(t)\Theta(\tau) \left[ \sin 4K\tau - \sin 4K(t+\tau) \right] \\
&+80 \, \Theta(t)\Theta(\tau) \left[ \sin K(2t - 4\tau) + 2\sin K(2t + 4\tau) - \sin 4K(t+\tau) \right] \\
&- 116 \, \Theta(t)\Theta(\tau) \left[ \sin 4K(t-\tau) + \sin 4K(t+\tau) \right].
\end{align}

In the presence of generic perturbations, the planons become dynamical and gain nonflat dispersion.
We focus on the processes for $\nu \in \mathbf{4}$ and $\nu \in \mathbf{6}$; $\nu \in \mathbf{4}$ process is important to identify deconfinement, and $\nu \in \mathbf{6}$ process distinguishes X-cube model from the toric code.
For the four-cube state $|\nu\rangle$, we can use the expression derived for the toric code, Eq.~(\ref{eq:chi3nu2_2d}) with $\epsilon_p = 2K + w(\cos p_x + \cos p_y)$.
Since the planons are restricted to a two-dimensional plane, nonlinear dynamics of two dispersive planons is equivalent to that of $2 \times 3L$ copies of toric codes.
The factor $3L$ comes from $3L$ distinct two dimensional planes of the cubic lattice, and the factor of 2 comes from two different ways to pair up four fractons (four edge sharing cubes excited by an action of $Z$ to the ground state. See the second figure in FIG.~\ref{fig:cross}) into two planons.
Hence, we only need to newly derive the expression for $|\nu\rangle \in \mathbf{6}$.

\begin{figure}[t]
\centering
\includegraphics[width=0.5\textwidth]{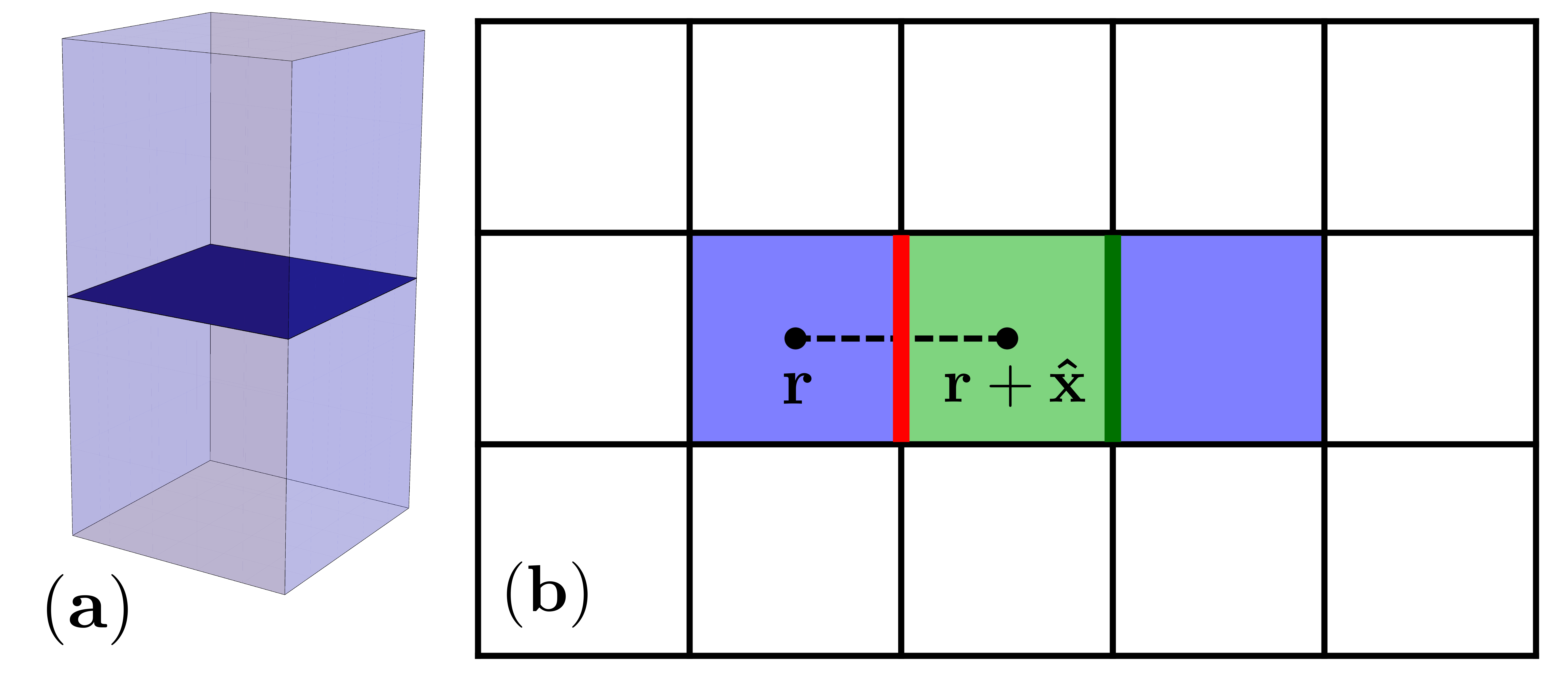}
\caption{(a) Location of a planon is labelled with centre of the face (marked as dark blue) shared by two neighbouring cubes. (b) Bird's eye view of FIG.~\ref{fig:fractons} (g). The planons and fractons are marked as blue and green squares, respectively. Note that the planons are not restricted to the same plane; one planon is located one ``floor'' higher than the other.
Immobile fractons are also blocking motions of both planons because they are occupying two separated floors.
The link can be uniquely specified with two neighbouring squares sharing the link (e.g., $\langle \mathbf{r}, \mathbf{r+\hat{x}}\rangle$) and which ``floor" those two squares belong to.}
\label{fig:birdseyeview}
\end{figure}

When there are six fractons, four fractons can be paired up to two planons moving in two-dimensional planes (FIG.~\ref{fig:fractons}). The remaining two fractons stay immobile.
The two planons can be either confined to the planes parallel [FIG.~\ref{fig:fractons} (g)] or perpendicular [FIG.~\ref{fig:fractons} (d, e, h)] to each other.
For simplicity, we focus on the case where two planons are moving in two parallel planes [FIG.~\ref{fig:fractons} (g)].
Then we can describe the motion of both planons in the two-dimensional plane (the bird's eye view to the two parallel planes) [FIG.~\ref{fig:birdseyeview} (b)].
Since a planon is a composite of two face-sharing cubes, let's label the location of the planon by the center of the face shared by both cubes [FIG.~\ref{fig:birdseyeview} (a)].
Note that two planons are not confined in the same plane.
If we think of the cubic lattice as a multilayered square lattices, one planon is moving in a plane one ``floor" higher than the other planon.
Hence, we will also label the planon with the floor index $n$.
Then we can label $Z$ operators acting on the red and green links in FIG.~\ref{fig:birdseyeview} (b) as
$Z_{\mathbf{r},\mathbf{r+\hat{x}}}^{(n)}$ and $Z_{\mathbf{r+\hat{x}},\mathbf{r+2\hat{x}}}^{(n+1)}$, respectively (if two planons are moving in the $n$ and $(n+1)$-th floors).

Then
\begin{align}
&\chi_z^{(a/b)}(\nu\in\mathbf{6};t,\tau) = \frac{2}{3L^3} \Theta(t)\Theta(\tau) \, \mathrm{Im}\left[
\frac{1}{2^2} \sum_{p_1 p_2}\sum_{q,\tilde{q},q'}\sum_{r_1 r_2} \mathcal{R}^{(a/b)}(p_1, p_2, q,\tilde{q},r_1,r_2;t,\tau) \right. \nonumber \\
&\left. \times \left( 3 \sum_{n=1}^L \sum_{\delta=\pm 1} \frac{1}{2^4}\sum_{jklm}\sum_{\eta_j\eta_k\eta_l\eta_m}
\langle 0 | Z_{j,j+\eta_j}^{(n)} |p_1,p_2\rangle \langle p_1, p_2 | Z_{k,k+\eta_k}^{(n+\delta)}|q,\tilde{q},q'\rangle\langle q,\tilde{q},q'| Z_{l,l+\eta_l}^{(n+\delta)} |r_1,r_2\rangle
\langle r_1,r_2 | Z_{m,m+\eta_m}^{(n)}|0\rangle \right.\right. \nonumber \\
&\left.\left. + 3 \sum_{n=1}^L \sum_{\delta=\pm 1} \frac{1}{2^4}\sum_{jklm}\sum_{\eta_j\eta_k\eta_l\eta_m}
\langle 0 | Z_{j,j+\eta_j}^{(n)} |p_1,p_2\rangle \langle p_1, p_2 | Z_{k,k+\eta_k}^{(n+\delta)}|q,\tilde{q},q'\rangle\langle q,\tilde{q},q'| Z_{l,l+\eta_l}^{(n)} |r_1,r_2\rangle
\langle r_1,r_2 | Z_{m,m+\eta_m}^{(n+\delta)}|0\rangle \right) \right],
\label{eq:chi_6_p1}
\end{align}
where $\eta_j, \eta_k, \eta_l, \eta_m =\pm \hat{x}, \pm\hat{y}$, and $q, \tilde{q}$ are the two-dimensional momenta of two planons.
Since the fractons are immobile, the resolution of identity with the real-space basis is physically more intuitive. However, we use the momentum basis also for fractons for simpler calculations. 
Recall that $|x\rangle = \frac{1}{L} \sum_{q'} e^{iq'\cdot x} |q'\rangle$; assigning momentum label $q'$ to the fracton does not imply mobility of fracton.
The mobility information is included in the Hamiltonian, which is relevant to the phase factor $\mathcal{R}^{(a/b)}$; the phase factor $\mathcal{R}^{(a/b)}$ are independent of $q'$ because the fractons are immobile,

Due to the momentum conservation imposed by the Kronecker delta in Eq.~(\ref{eq:0Zp}), we can simplify the notation $\mathcal{R}^{(a/b)}(p,q,\tilde{q},r;t,\tau) \equiv \mathcal{R}^{(a/b)}(p, -p, q, \tilde{q}, r, -r;t,\tau)$ with $p_1 \equiv p$, $p_2 = -p$, and $r_1 \equiv r$, $r_2 = -r$.
Also product of the matrix elements inside the square bracket of Eq.~(\ref{eq:chi_6_p1}) is independent of the floor/layer index $n$ and $\delta = \pm 1$.
For concise notation, let's write $Z_{j,j+\eta_j} \equiv Z_{j,j+\eta_j}^{(n)}$ and $\widetilde{Z}_{j,j+\eta_j}^{(\pm)} \equiv Z_{j,j+\eta_j}^{(n\pm1)}$.
Then
\begin{align}
&\chi_z^{(a/b)}(\nu\in\mathbf{6};t,\tau) = \frac{1}{L^2} \Theta(t)\Theta(\tau) \, \mathrm{Im}\left[
\sum_{p, r}\sum_{q,\tilde{q},q'} \mathcal{R}^{(a/b)}(p, q,\tilde{q},r;t,\tau) \right. \nonumber \\
&\left. \times \left (\frac{1}{2^4}\sum_{jklm}\sum_{\eta_j\eta_k\eta_l\eta_m}
\langle 0 | Z_{j,j+\eta_j} |p,-p\rangle \langle p, -p | \widetilde{Z}_{k,k+\eta_k}^{(+)}|q,\tilde{q},q'\rangle\langle q,\tilde{q},q'| \widetilde{Z}_{l,l+\eta_l}^{(+)} |r,-r\rangle
\langle r,-r | Z_{m,m+\eta_m}|0\rangle \right. \right. \label{eq:chi_6_abba} \\
&\left. \left. + \frac{1}{2^4}\sum_{jklm}\sum_{\eta_j\eta_k\eta_l\eta_m}
\langle 0 | Z_{j,j+\eta_j} |p,-p\rangle \langle p, -p| \widetilde{Z}_{k,k+\eta_k}^{(+)}|q,\tilde{q},q'\rangle\langle q,\tilde{q},q'| Z_{l,l+\eta_l} |r,-r\rangle
\langle r,-r | \widetilde{Z}_{m,m+\eta_m}^{(+)}|0\rangle \right) \right].
\label{eq:chi_6_abab}
\end{align}

To calculate the matrix element,
\begin{align}
\frac{1}{2}\sum_{k,\eta_k}\langle p, -p | \widetilde{Z}_{k,k+\eta_k}^{(+)}|q,\tilde{q},q'\rangle
& = \frac{1}{2L^5} \sum_{k,\eta_k} \sideset{}{'}\sum_{\xi_1 \xi_2} \sideset{}{'}\sum_{\zeta \tilde{\zeta} \zeta'} e^{-i p \cdot(\xi_1 - \xi_2)}e^{i q \cdot \zeta + i \tilde{q} \cdot \tilde{\zeta} + i q' \cdot \zeta' } \langle \xi_1, \xi_2| \widetilde{Z}^{(+)}_{k,k+\eta_k}|\zeta,\tilde{\zeta},\zeta'\rangle,
\end{align}
we need to calculate $\langle \xi_1, \xi_2| \widetilde{Z}^{(+)}_{k,k+\eta_k}|\zeta,\widetilde{\zeta},\zeta'\rangle$. Here $\xi_1, \xi_2$ are the positions of two planons having momentum $p$ and $-p$, and $\zeta$, $\widetilde{\zeta}$ are the positions of two planons having momentum $q$ and $\tilde{q}$. $\zeta'$ is the position of the immobile fractons.
The restricted sum excludes the cases where planons and fractons are overlapping, \textit{i.e.}, the planons and fractons are treated as hardcore bosons.

Note that $\widetilde{Z}_{k,k+\eta_k}^{(+)}$ creates/annihilates two planons living in the $(n+1)$-th layer of the cubic lattice. Within the layer, they are located at $\mathbf{R}_k$ and $\mathbf{R}_k+\eta_k$.
So we can identify
\begin{align}
\widetilde{Z}_{k,k+\eta_k}^{(+)} \to \tilde{a}_k \tilde{a}_{k+\eta_k}.
\end{align}
When a planon at the $n$-th floor overlaps with a planon at the $(n\pm1)$-th floor, then we get two immobile fractons. Therefore we can identify
\begin{align}
a_j \tilde{a}_j \to f_j,
\end{align}
where $f_j$ is annihilation operator of two immobile fractons. These two immobile fractons are located at the same location $\mathbf{R}_j$ in the two-dimensional plane but belong to different layers [FIG.~\ref{fig:fractons} (g)].
Therefore
\begin{align}
\frac{1}{2}\sum_{k} \langle \xi_1, \xi_2| \widetilde{Z}^{(+)}_{k,k+\eta_k}|\zeta,\widetilde{\zeta},\zeta'\rangle
&\to \frac{1}{2}\sum_{k}\langle 0 | a_{\xi_1} a_{\xi_2} \tilde{a}_k \tilde{a}_{k+\eta_k} a_\zeta^\dagger a_{\tilde{\zeta}}^\dagger f_{\zeta'}^\dagger |0\rangle \\
&\to \frac{1}{2}\sum_{k} \delta_{\xi_1,k}\langle 0| a_{\xi_2} \tilde{a}_{k+\eta_k} f_k a_\zeta^\dagger a_{\tilde{\zeta}}^\dagger f_{\zeta'}^\dagger|0\rangle
+ \delta_{\xi_2,k}\langle 0| a_{\xi_1} \tilde{a}_{k+\eta_k} f_k a_\zeta^\dagger a_{\tilde{\zeta}}^\dagger f_{\zeta'}^\dagger|0\rangle \\
&+ \frac{1}{2}\sum_{k} \delta_{\xi_1,k+\eta_k}\langle 0| a_{\xi_2} \tilde{a}_{k} f_k a_\zeta^\dagger a_{\tilde{\zeta}}^\dagger f_{\zeta'}^\dagger|0\rangle
+ \delta_{\xi_2,k+\eta_k}\langle 0| a_{\xi_1} \tilde{a}_{k} f_k a_\zeta^\dagger a_{\tilde{\zeta}}^\dagger f_{\zeta'}^\dagger|0\rangle \\
&= \sum_k (\delta_{\xi_1,k}\delta_{\xi_2,\zeta} + \delta_{\xi_1,\zeta}\delta_{\xi_2,k} )
\delta_{\tilde{\zeta},k+\eta_k}\delta_{\zeta',k},
\end{align}
so that
\begin{align}
&\frac{1}{2}\sum_{k,\eta_k}\langle p, -p | \widetilde{Z}_{k,k+\eta_k}^{(+)}|q,\tilde{q},q'\rangle
= \frac{1}{L^5}\sum_{k,\eta_k}  \sideset{}{'}\sum_{\xi_1 \xi_2} \sideset{}{'}\sum_{\zeta \tilde{\zeta} \zeta'} e^{-i p \cdot(\xi_1 - \xi_2)}e^{i q \cdot \zeta + i \tilde{q} \cdot \tilde{\zeta} + i q' \cdot \zeta' } 
(\delta_{\xi_1,k}\delta_{\xi_2,\zeta} + \delta_{\xi_1,\zeta}\delta_{\xi_2,k} )
\delta_{\tilde{\zeta},k+\eta_k}\delta_{\zeta',k} \nonumber \\
&= \frac{1}{L^5} \sum_{k,\eta_k}\sum_{\zeta\neq k,k+\eta_k}
\left(e^{-ip\cdot(R_k -\zeta)}+e^{-ip\cdot(\zeta-R_k)}  \right)
e^{i q \cdot \zeta + i(\tilde{q}+q')\cdot R_k + i\tilde{q} \cdot \eta_k} \\
&= \frac{1}{L^5} \sum_{k,\eta_k}\sum_{\zeta} e^{i\tilde{q}\cdot \eta_k}
\left( e^{i(\tilde{q}+q'-p)\cdot R_k}e^{i(q+p)\cdot\zeta} + e^{i(\tilde{q}+q'+p)\cdot R_k}e^{i(q-p)\cdot\zeta}\right)
-\frac{1}{L^5} \sum_{k,\eta_k} e^{i(q+\tilde{q}+q')\cdot R_k}e^{i\tilde{q}\cdot \eta_k}
\left( 2 + e^{i(q+p)\cdot \eta_k} + e^{i(q-p)\cdot \eta_k} \right) \nonumber\\
&= \frac{1}{L} \sum_{\eta_k} e^{i\tilde{q}\cdot \eta_k}
\left(\delta_{\tilde{q}+q',p}\delta_{q,-p} + \delta_{\tilde{q}+q',-p}\delta_{q,p} \right)
-\frac{1}{L^3} \delta_{q+\tilde{q}+q',0} \sum_{\eta_k} 2 e^{i\tilde{q}\cdot \eta_k} + e^{i(q+\tilde{q}+p)\cdot \eta_k}+e^{i(q+\tilde{q}-p)\cdot\eta_k} \\
&=\frac{2}{L}(\cos \tilde{q}_x + \cos \tilde{q}_y) \left(\delta_{\tilde{q}+q',p}\delta_{q,-p} + \delta_{\tilde{q}+q',-p}\delta_{q,p} \right) \nonumber \\
&-\frac{2}{L^3} \left[
2 \cos \tilde{q}_x + 2\cos \tilde{q}_y + \cos(q_x + \tilde{q}_x +p_x) + \cos(q_y + \tilde{q}_y +p_y)
+ \cos(q_x + \tilde{q}_x -p_x) + \cos(q_y + \tilde{q}_y -p_y)
\right]
\end{align}

With similar calculations, we can get the expression for the third-order susceptibilities
\begin{equation}
\chi_z^{(a/b)}(\nu\in\mathbf{6};t,\tau) = \frac{8}{L^4}\,\Theta(t)\Theta(\tau)
\left[ \mathcal{A}(t,\tau) + \mathcal{B}(t,\tau) \right],
\end{equation}
where
\begin{align}
&\mathcal{A}(t,\tau) = \sum_{p,\tilde{q}} \mathrm{Im} \left[
\mathcal{R}^{(a/b)}(p,p,\tilde{q},p;t,\tau)
+\mathcal{R}^{(a/b)}(p,-p,\tilde{q},p;t,\tau)
+\mathcal{R}^{(a/b)}(p,p,\tilde{q},-p;t,\tau)+\mathcal{R}^{(a/b)}(p,-p,\tilde{q},-p;t,\tau)\right] \nonumber\\
&\qquad\qquad\qquad \qquad
\times(\cos p_x +\cos p_y)^2(\cos \tilde{q}_x + \cos \tilde{q}_y)^2 \\
&- \frac{1}{L^2} \sum_{p,\tilde{q},r}
\mathrm{Im}\left[\mathcal{R}^{(a/b)}(p,p,\tilde{q},r;t,\tau)\right]
(\cos p_x + \cos p_y) (\cos r_x + \cos r_y) (\cos \tilde{q}_x + \cos \tilde{q}_y) \nonumber \\
&\times \left[2 \cos \tilde{q}_x + 2 \cos \tilde{q}_y + \cos(\tilde{q}_x + p_x + r_x) + \cos(\tilde{q}_y + p_y + r_y)+\cos(\tilde{q}_x + p_x - r_x) + \cos(\tilde{q}_y + p_y - r_y)\right] \\
&- \frac{1}{L^2} \sum_{p,\tilde{q},r}
\mathrm{Im}\left[\mathcal{R}^{(a/b)}(p,-p,\tilde{q},r;t,\tau)\right]
(\cos p_x + \cos p_y) (\cos r_x + \cos r_y) (\cos \tilde{q}_x + \cos \tilde{q}_y) \nonumber \\
&\times \left[2 \cos \tilde{q}_x + 2 \cos \tilde{q}_y + \cos(\tilde{q}_x - p_x + r_x) + \cos(\tilde{q}_y - p_y + r_y)+\cos(\tilde{q}_x - p_x - r_x) + \cos(\tilde{q}_y - p_y - r_y)\right] \\
&- \frac{1}{L^2} \sum_{p,\tilde{q},r}
\mathrm{Im}\left[\mathcal{R}^{(a/b)}(p,r,\tilde{q},r;t,\tau)\right]
(\cos p_x + \cos p_y) (\cos r_x + \cos r_y) (\cos \tilde{q}_x + \cos \tilde{q}_y) \nonumber \\
&\times \left[2 \cos \tilde{q}_x + 2 \cos \tilde{q}_y + \cos(\tilde{q}_x + p_x + r_x) + \cos(\tilde{q}_y + p_y + r_y)+\cos(\tilde{q}_x - p_x + r_x) + \cos(\tilde{q}_y - p_y + r_y)\right] \\
&- \frac{1}{L^2} \sum_{p,\tilde{q},r}
\mathrm{Im}\left[\mathcal{R}^{(a/b)}(p,-r,\tilde{q},r;t,\tau)\right]
(\cos p_x + \cos p_y) (\cos r_x + \cos r_y) (\cos \tilde{q}_x + \cos \tilde{q}_y) \nonumber \\
&\times \left[2 \cos \tilde{q}_x + 2 \cos \tilde{q}_y + \cos(\tilde{q}_x + p_x - r_x) + \cos(\tilde{q}_y + p_y - r_y)+\cos(\tilde{q}_x - p_x - r_x) + \cos(\tilde{q}_y - p_y - r_y)\right]\\
&+\frac{1}{L^4} \sum_{p,q,\tilde{q},r}
\mathrm{Im}\left[ \mathcal{R}^{(a/b)}(p,q,\tilde{q},r;t,\tau)\right]
(\cos p_x + \cos p_y ) (\cos r_x +\cos r_y) \nonumber \\
&\times \left[2 \cos \tilde{q}_x + 2 \cos \tilde{q}_y + \cos(q_x + \tilde{q}_x + p_x)
+ \cos(q_y + \tilde{q}_y + p_y)+\cos(q_x + \tilde{q}_x - p_x) + \cos(q_y + \tilde{q}_y - p_y)\right] \nonumber \\
&\times \left[2 \cos \tilde{q}_x + 2 \cos \tilde{q}_y + \cos(q_x + \tilde{q}_x + r_x)
+ \cos(q_y + \tilde{q}_y + r_y)+\cos(q_x + \tilde{q}_x - r_x) + \cos(q_y + \tilde{q}_y - r_y)\right],
\end{align}

\begin{align}
&\mathcal{B}(t,\tau) = \sum_{p,r} \mathrm{Im} \left[
\mathcal{R}^{(a/b)}(p,p,r,r;t,\tau)
+\mathcal{R}^{(a/b)}(p,p,-r,r;t,\tau)
+\mathcal{R}^{(a/b)}(p,-p,r,r;t,\tau)+\mathcal{R}^{(a/b)}(p,-p,-r,r;t,\tau)\right] \nonumber\\
&\qquad\qquad\qquad \qquad
\times(\cos p_x +\cos p_y)^2(\cos r_x + \cos r_y)^2 \\
&- \frac{1}{L^2} \sum_{p,\tilde{q},r}
\mathrm{Im}\left[\mathcal{R}^{(a/b)}(p,p,\tilde{q},r;t,\tau)\right]
(\cos p_x + \cos p_y) (\cos r_x + \cos r_y) (\cos \tilde{q}_x + \cos \tilde{q}_y) \nonumber \\
&\times \left[2 \cos p_x + 2 \cos p_y + \cos(\tilde{q}_x + p_x + r_x) + \cos(\tilde{q}_y + p_y + r_y)+\cos(\tilde{q}_x + p_x - r_x) + \cos(\tilde{q}_y + p_y - r_y)\right] \\
&- \frac{1}{L^2} \sum_{p,\tilde{q},r}
\mathrm{Im}\left[\mathcal{R}^{(a/b)}(p,-p,\tilde{q},r;t,\tau)\right]
(\cos p_x + \cos p_y) (\cos r_x + \cos r_y) (\cos \tilde{q}_x + \cos \tilde{q}_y) \nonumber \\
&\times \left[2 \cos p_x + 2 \cos p_y + \cos(\tilde{q}_x - p_x + r_x) + \cos(\tilde{q}_y - p_y + r_y)+\cos(\tilde{q}_x - p_x - r_x) + \cos(\tilde{q}_y - p_y - r_y)\right] \\
&- \frac{1}{L^2} \sum_{p,q,r}
\mathrm{Im}\left[\mathcal{R}^{(a/b)}(p,q,r,r;t,\tau)\right]
(\cos p_x + \cos p_y) (\cos r_x + \cos r_y) (\cos q_x + \cos q_y) \nonumber \\
&\times \left[2 \cos r_x + 2 \cos r_y + \cos(q_x + p_x + r_x) + \cos(q_y + p_y + r_y)+\cos(q_x - p_x + r_x) + \cos(q_y - p_y + r_y)\right] \\
&- \frac{1}{L^2} \sum_{p,q,r}
\mathrm{Im}\left[\mathcal{R}^{(a/b)}(p,q,-r,r;t,\tau)\right]
(\cos p_x + \cos p_y) (\cos r_x + \cos r_y) (\cos q_x + \cos q_y) \nonumber \\
&\times \left[2 \cos r_x + 2 \cos r_y + \cos(q_x + p_x - r_x) + \cos(q_y + p_y - r_y)+\cos(q_x - p_x - r_x) + \cos(q_y - p_y - r_y)\right]\\
&+\frac{1}{L^4} \sum_{p,q,\tilde{q},r}
\mathrm{Im}\left[ \mathcal{R}^{(a/b)}(p,q,\tilde{q},r;t,\tau)\right]
(\cos p_x + \cos p_y ) (\cos r_x +\cos r_y) \nonumber \\
&\times \left[2 \cos \tilde{q}_x + 2 \cos \tilde{q}_y + \cos(q_x + \tilde{q}_x + p_x)
+ \cos(q_y + \tilde{q}_y + p_y)+\cos(q_x + \tilde{q}_x - p_x) + \cos(q_y + \tilde{q}_y - p_y)\right] \nonumber \\
&\times \left[2 \cos q_x + 2 \cos q_y + \cos(q_x + \tilde{q}_x + r_x)
+ \cos(q_y + \tilde{q}_y + r_y)+\cos(q_x + \tilde{q}_x - r_x) + \cos(q_y + \tilde{q}_y - r_y)\right],
\end{align}
with
\begin{align}
\mathcal{R}^{(a)}(p,q,\tilde{q},r;t,\tau) &= 2 e^{2i(\varepsilon_p-\varepsilon_r)\tau}e^{i(\varepsilon_q + \varepsilon_{\tilde{q}} + 2K - 2\varepsilon_r)t}
+ e^{i (\varepsilon_q + \varepsilon_{\tilde{q}} + 2K)\tau}e^{i(\varepsilon_q + \varepsilon_{\tilde{q}} + 2K-2\varepsilon_r)t}+e^{-i(\varepsilon_q + \varepsilon_{\tilde{q}} + 2K)\tau}e^{-2i\varepsilon_p t},\\
\mathcal{R}^{(b)}(p,q,\tilde{q},r;t,\tau) &= e^{-2i \varepsilon_r\tau}e^{i(\varepsilon_q + \varepsilon_{\tilde{q}} + 2K - 2\varepsilon_r)t}
+ 2e^{2i\varepsilon_p\tau}e^{i(\varepsilon_q + \varepsilon_{\tilde{q}} + 2K-2\varepsilon_r)t} + e^{-2i\varepsilon_r\tau}e^{-2i\varepsilon_p t}.
\end{align}
The dispersion for a single planon $\varepsilon_s = 2K + w(\cos s_x + \cos s_y)$ for $s = p, q, \tilde{q}, r$.
Above expressions reproduce the exact result,
$\chi_z^{(a/b)}(\nu\in\mathbf{6};t,\tau) = 80 \,\mathrm{Im}\,\mathcal{R}^{(a/b)}(\mathbf{4},\mathbf{6},\mathbf{4};t,\tau)$ in Eq.~(\ref{eq:chi_xcube_z}), when $w = 0$.

\section{One-sided Fourier transformation}

After we calculate the nonlinear susceptibilities $\chi(t,\tau)$ as a function of two times, we use the two-dimensional Fourier transformation $\mathcal{F}_{t,\tau}$ to obtain the two-dimensional nonlinear spectrum $\tilde{\chi}(\omega_t,\omega_\tau) = \mathcal{F}_{t,\tau} [\chi(t,\tau)]$.
Due to the causality, the time $t$ and $\tau$ cannot be negative; we cannot think of the second pulse before the first pulse or the nonlinear response of matter before the pulse stimulates the system.
Hence, the nonlinear susceptibilities [Eqs.~(\ref{chi31}) and (\ref{chi32})] include the Heaviside step functions to make sure that both times are always positive.
So the Fourier transformation of the susceptibilities is essentially one-sided, \textit{i.e.}, the Fourier integral is applied only along the positive time direction.
In this Appendix, we discuss the consequence of this one-sided nature of Fourier transformation in order to discriminate more interesting and intrinsic information about the system from the Fourier transformed susceptibilities.

Since the general forms of the third-order susceptibilities in Eqs.~(\ref{chi31}) and (\ref{chi32}) are too complex, let's consider a simplified expression for better understanding:
\begin{align}
\chi(t,\tau) = i \Theta(t)\Theta(\tau) \sum_{a,b} S_{ab}e^{-i \lambda_{ab} t}e^{-i \tilde{\lambda}_{ab}\tau}.
\end{align}
Then the Fourier transformation with respect to $t$ and $\tau$ is
\begin{align}
\tilde{\chi}(\omega_t,\omega_\tau) = \sum_{a,b} i S_{ab}
\left[\frac{1}{2\pi} \int_{-\infty}^{\infty}\Theta(t) e^{-i \lambda_{ab} t} e^{i \omega_t t} \,dt\right]
\left[\frac{1}{2\pi} \int_{-\infty}^{\infty}\Theta(\tau) e^{-i \tilde{\lambda}_{ab} \tau} e^{i \omega_\tau \tau} \,d\tau \right].
\end{align}
By the convolution theorem,
\begin{align}
\frac{1}{2\pi} \int_{-\infty}^{\infty}\Theta(t) e^{-i \lambda_{ab} t} e^{i \omega_t t} \,dt
&= \int_{-\infty}^{\infty}
\left[\frac{1}{2\pi}\int_{-\infty}^{\infty} \Theta(u) e^{i x u} \,du \right]
\left[\frac{1}{2\pi}\int_{-\infty}^{\infty} e^{-i \lambda_{ab} v}  e^{i (\omega_t - x) v} \,dv \right] \,dx \\
&=\int_{-\infty}^{\infty}
\frac{1}{2} \left[ \delta(x) + \frac{1}{i\pi} \mathcal{P}\left(\frac{1}{x} \right) \right]
\delta(\omega_t-\lambda_{ab}-x) \,dx \\
&=\frac{1}{2} \delta(\omega_t - \lambda_{ab}) + \frac{1}{2\pi i} \mathcal{P}\left(\frac{1}{\omega_t - \lambda_{ab}} \right).
\end{align}
Therefore
\begin{align}
\tilde{\chi}(\omega_t, \omega_\tau) &= \sum_{a,b} i S_{ab}
\left[\frac{1}{2} \delta(\omega_t - \lambda_{ab}) + \frac{1}{2\pi i} \mathcal{P}\left(\frac{1}{\omega_t - \lambda_{ab}} \right) \right]
\left[\frac{1}{2} \delta(\omega_\tau - \tilde{\lambda}_{ab}) + \frac{1}{2\pi i} \mathcal{P}\left(\frac{1}{\omega_\tau - \tilde{\lambda}_{ab}} \right) \right] \\
&=\frac{1}{4} \sum_{a,b} S_{ab} \left \{
\frac{1}{\pi} \left[
\delta(\omega_t - \lambda_{ab})\mathcal{P}\left(\frac{1}{\omega_\tau - \tilde{\lambda}_{ab}} \right)
+ \delta(\omega_\tau -\tilde{\lambda}_{ab})\mathcal{P}\left(\frac{1}{\omega_t - \lambda_{ab}} \right)
\right]\right. \\
&\qquad\qquad\qquad\left.+ i \left[
\delta(\omega_t - \lambda_{ab}) \delta(\omega_\tau - \tilde{\lambda}_{ab})
- \frac{1}{\pi^2} \mathcal{P}\left(\frac{1}{(\omega_t -\lambda_{ab})(\omega_\tau -\tilde{\lambda}_{ab}} \right)
\right]\right\}
\end{align}

If we take the imaginary part of the Fourier spectrum,
\begin{align}
\mathrm{Im}\left[\tilde{\chi}(\omega_t,\omega_\tau)\right]
= \frac{1}{4} \sum_{a,b} S_{ab}
\left[
\delta(\omega_t - \lambda_{ab}) \delta(\omega_\tau - \tilde{\lambda}_{ab})
- \frac{1}{\pi^2} \mathcal{P}\left(\frac{1}{(\omega_t - \lambda_{ab})(\omega_\tau - \tilde{\lambda}_{ab})} \right)
\right],
\end{align}
then the first product of the delta functions provide information how two energy levels $|a\rangle$ and $|b\rangle$ are correlatetd.
The principal values are responsible for the horizontal and vertical line signals in the two-dimensional nonlinear spectrum.
Those lines are asymptotes of $1/(\omega_t -\lambda_{ab})$ and $1/(\omega_\tau - \tilde{\lambda}_{ab})$.
\end{document}